\documentclass[acmsmall,screen]{acmart}

\usepackage[linesnumbered,commentsnumbered,ruled,noend]{algorithm2e}
\usepackage{multirow}

\AtBeginDocument{%
  }

\setcopyright{acmcopyright}
\copyrightyear{2018}
\acmYear{2018}
\acmDOI{XXXXXXX.XXXXXXX}

\acmJournal{JACM}
\acmVolume{37}
\acmNumber{4}
\acmArticle{111}
\acmMonth{8}

\begin{document}

\title{MultiGraphMatch: a subgraph matching algorithm for multigraphs}


\author{Giovanni Micale}
\email{giovanni.micale@unict.it}
\orcid{0000-0002-4953-026X}
\affiliation{%
  \institution{University of Catania}
  %\streetaddress{P.O. Box 1212}
  \city{Catania}
  \country{Italy}
}

\author{Antonio Di Maria}
\email{antoniodm@unict.it}
\orcid{0009-0008-9300-2297}
\affiliation{%
  \institution{University of Catania}
  %\streetaddress{1 Th{\o}rv{\"a}ld Circle}
  \city{Catania}
  \country{Italy}
}

\author{Roberto Grasso}
\email{roberto.grasso@dfa.unict.it}
\orcid{0000-0002-7641-1979}
\affiliation{%
  \institution{University of Catania}
  \city{Catania}
  \country{Italy}
}

\author{Vincenzo Bonnici}
\email{vincenzo.bonnici@unipr.it}
\orcid{0000-0002-1637-7545}
\affiliation{%
  \institution{University of Parma}
  %\streetaddress{ì}
  \city{Parma}
  %\state{Texas}
  \country{Italy}
  %\postcode{43124}
}

\author{Alfredo Ferro}
\email{alfredo.ferro@unict.it}
\orcid{0000-0002-9431-5788}
\affiliation{%
 \institution{University of Catania}
 %\streetaddress{Rono-Hills}
 \city{Catania}
 %\state{New York}
 \country{Italy}
}

\author{Dennis Shasha}
\email{shasha@cs.nyu.edu}
\orcid{0000-0002-7036-3312}
\affiliation{%
 \institution{New York University}
 %\streetaddress{Rono-Hills}
 \city{New York}
 \state{New York}
 \country{USA}
}

\author{Rosalba Giugno}
\email{rosalba.giugno@univr.it}
\orcid{0000-0001-9843-7638}
\affiliation{%
  \institution{University of Verona}
  %\streetaddress{8600 Datapoint Drive}
  \city{Verona}
  %\state{Texas}
  \country{Italy}
  %\postcode{78229}
}

\author{Alfredo Pulvirenti}
\email{alfredo.pulvirenti@unict.it}
\orcid{0000-0002-9764-0295}
\affiliation{%
  \institution{University of Catania}
  %\streetaddress{30 Shuangqing Rd}
  \city{Catania}
  %\state{Beijing Shi}
  \country{Italy}
}

\renewcommand{\shortauthors}{Micale et al.}


\begin{abstract}

Subgraph matching is the problem of finding all the occurrences of a small graph, called the query, in a larger graph, called the target. Although the problem has been widely studied in simple graphs, few solutions have been proposed for multigraphs, in which two nodes can be connected by multiple edges, each denoting a possibly different type of relationship. In our new algorithm MultiGraphMatch, nodes and edges can be associated with labels and multiple properties. MultiGraphMatch introduces a novel data structure called bit matrix to efficiently index both the query and the target and filter the set of target edges that are matchable with each query edge. In addition, the algorithm proposes a new technique for ordering the processing of query edges based on the cardinalities of the sets of matchable edges. Using the CYPHER query definition language, MultiGraphMatch can perform queries with logical conditions on node and edge labels. We compare MultiGraphMatch with SuMGra and graph database systems Memgraph and Neo4J, showing comparable or better performance in all queries on a wide variety of synthetic and real-world graphs.

\end{abstract}





\received{20 February 2007}
\received[revised]{12 March 2009}
\received[accepted]{5 June 2009}

\maketitle

\section{Introduction}
\label{introduction} 
Graphs (also called networks) are mathematical objects that are used to model complex systems consisting of several entities that interact with each other in different ways. In a graph, entities are represented by nodes, whereas edges denote physical or virtual relationships between objects. 

Examples of graphs include social networks like Facebook or Instagram, where nodes represent people and edges indicate relationships among them, biological networks such as protein-protein interaction networks, where nodes are proteins and edges denote physical or chemical interactions between proteins, and infrastructural networks, such as the electrical grid, in which nodes are power stations and generators, and edges represent transmission lines. 

Graphs can be labeled, i.e. both nodes and edges can be associated with labels, denoting one or more classes to which a node belongs and the type of relationship between two nodes. For instance, in a social network, people can be labeled with the community to which they belong, and relationship edges can be labeled with "friend", "co-worker", "customer", etc. Note that these edges can be symmetric (e.g. as colleagues) as well as asymmetric (e.g. one is fan of the other). Thus, some edges may be directed and some bidirectional.

Graphs (labeled or not) can also be attributed, i.e. nodes and edges can be annotated with properties and corresponding values. For example, on a social network, a person can be associated with properties such as age, sex, or income. While labels denote fairly small classes or categories to which nodes or edges belong, properties can take values from unbounded domains, e.g. all integers.

In this paper, we will focus on labeled attributed multigraphs (simply denoted as multigraphs), i.e. labeled and attributed graphs where multiple directed or bidirectional edges between two nodes may exist. 

\begin{figure}
\centering
\includegraphics[width=12cm]{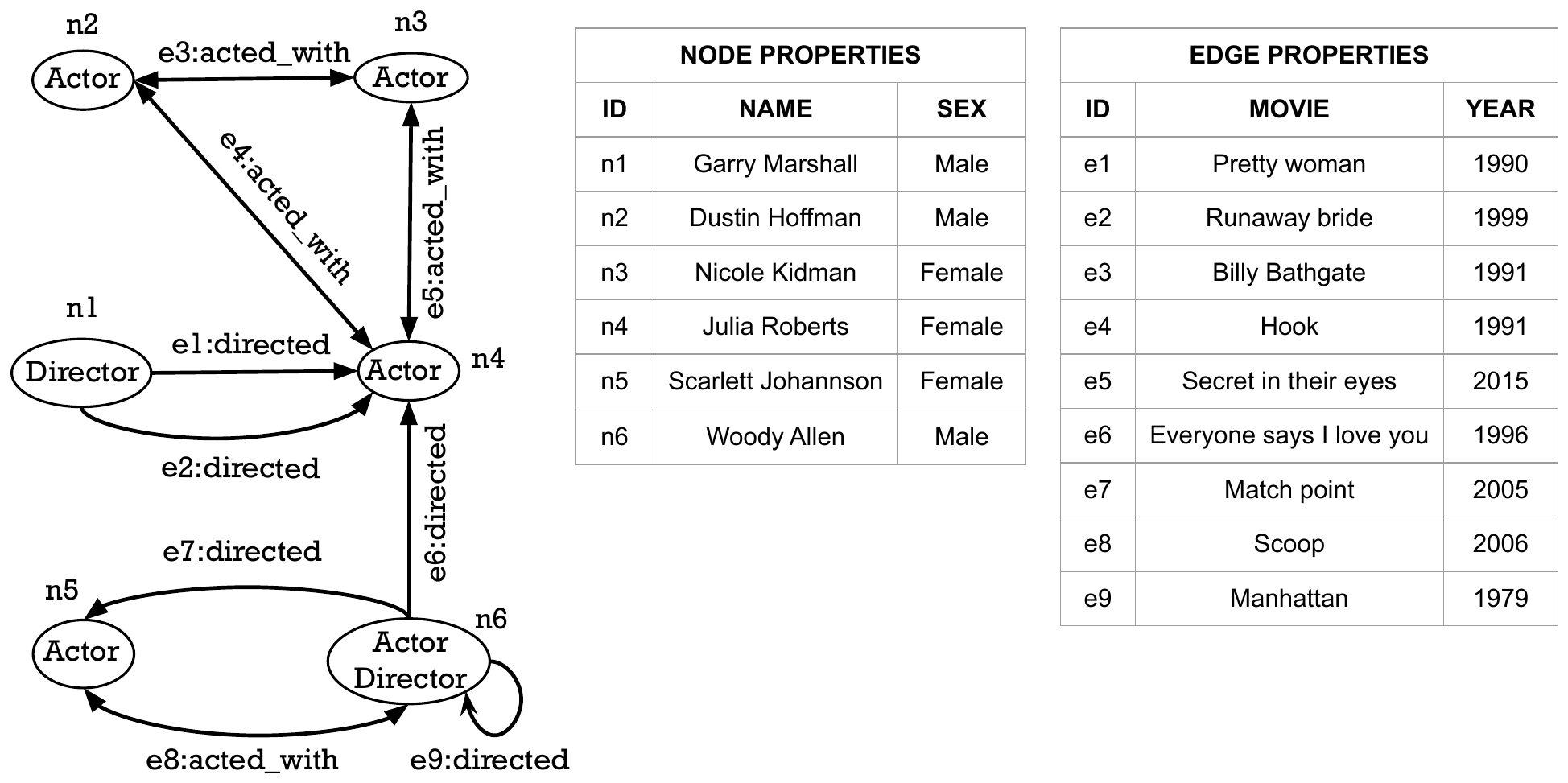}
  \caption{Example of multigraph with actors and directors of the movie industry.}
  \label{Multigraph_Example}
  \Description[Multigraph example]{Example of multigraph with actors and directors of the movie industry}
\end{figure}

An example of a multigraph is depicted in Fig. \ref{Multigraph_Example}, where nodes are actors and directors, two actors are linked by bidirectional edges iff they acted together in a movie and a directed edge connects a director $D$ to an actor $A$ iff $D$ directed $A$ in a movie.

Node labels represent the main professions of a movie star (`actor' and/or `director'), while edge labels denote the type of relationship between two nodes (`directed' or `acted\_with'). Node and edge properties represent, respectively, the name and sex of a movie star and the name and year of production of a movie referenced by a specific edge. 

One of the most common problems in graph analysis is Subgraph Matching (SM), which consists in finding all the occurrences of a small graph (called the query) in a larger graph (called the target). Matching implies that the query must be structurally equivalent to a subgraph of the target, and the node and edge labels of the query and the subgraph must be the same. \textcolor{black}{In the SM problem, there is at most one edge between any two nodes and each edge has at most one label.}

A generalization of the SM problem to multigraphs is the Sub-Multigraph Matching (SMM), which aims to identify all the occurrences of a query multigraph in a target multigraph.

%\color{red} 

To our knowledge, no algorithm designed to solve the SM problem has been applied directly when multiple labels are present on the edges. Adapting an SM algorithm to address the SMM problem would first require finding all possible correct mappings between query and target nodes. Each identified node mapping corresponds to a match, and thus a candidate solution to the problem. As a post-processing step, the algorithm must retrieve all possible correct edge mappings for a given candidate match $M$, trying to pair each edge connecting two query nodes $u$ and $v$ with each edge linking target nodes mapped to $u$ and $v$. Post-processing of matches is necessary, even when the algorithm uses data structures to filter candidates based on edge labels. Thus, filtering can help reduce the number of candidate matches, but the latter still need to be post-processed to find correct edge mappings.

\begin{figure}
\centering
\includegraphics[width=10cm]{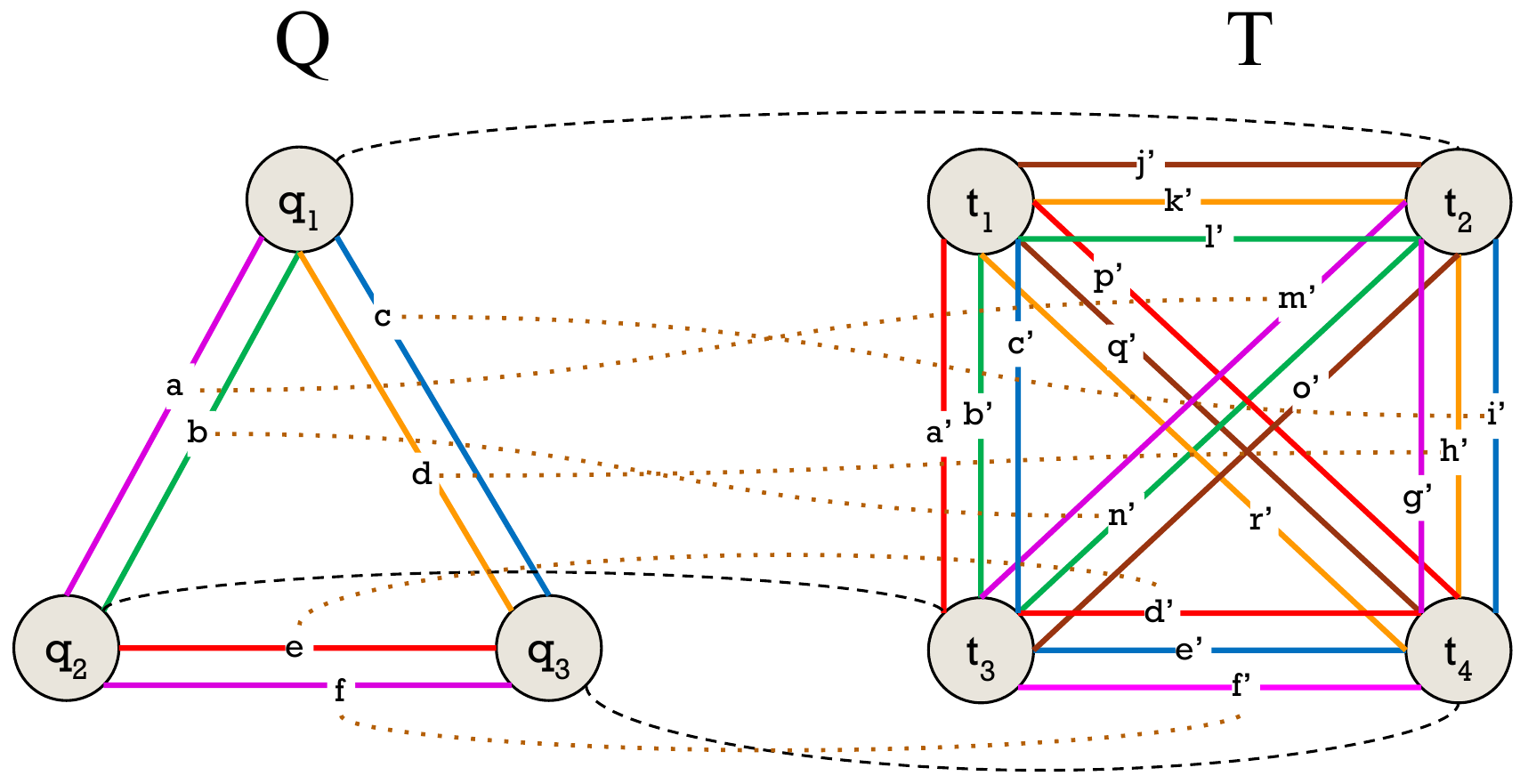}
  \caption{Example of sub-multigraph matching. There is only one occurrence of query $Q$ in target $T$. A subgraph matching algorithm for simple graphs needs to find all possible correct mappings between nodes of $Q$ and $T$, then post-process the results to find the correct mapping between query and target edge labels. By contrast, MultiGraphMatch can retrieve the correct node and edge mappings during the matching, without the need of post-processing.}
  \label{Example_Exponential}
  \Description[Exponential example]{Example of submultigraph matching to show exponential complexity of subgraph matching algorithms working on simple graphs}
\end{figure}

As an example, consider Fig. \ref{Example_Exponential}, where the query triangle $Q$ occurs only once in the target 4-clique $T$.
A subgraph matching algorithm operating on simple graphs would generate all possible $\binom{4}{3}$ occurrences of $Q$ in $T$, resulting in multiple candidates. By filtering node candidates based on edge labels, the algorithm could generate only one match with nodes $q_1$, $q_2$ and $q_3$ mapped to nodes $t_2$, $t_3$ and $t_4$, respectively. However, retrieving all possible correct edge mappings for the given candidate match implies considering $\binom{3}{2}$ pairings between query and target edges, for each pair of connected query nodes. If $E_Q$ and $E_T$ are the number of query and target edges, respectively, in the worst case post-processing of a candidate match requires $\binom{E_T}{E_Q}$, thus exponential in $E_T$ in the worst case.

%\color{black}

For these reasons, in the last few years, new algorithms have been specifically designed for the SMM problem, providing full or partial support for querying multigraphs.

Some solutions include graph database systems \cite{timon2021overview}, while others consist of in-memory algorithms \cite{anwar2022subgraph,ingalalli2016sumgra,moorman2021subgraph,sun2021subgraph}. Details of the related work are reported in Section \ref{related}.

%\color{red}

In contrast with classical subgraph matching techniques, algorithms designed specifically for multigraphs can find the correct mapping between query and target edges during matching, so they can directly identify the final set of solutions without requiring post-processing.

The main contributions of this paper are: 
\begin{itemize}
\item A new in-memory algorithm for the sub-multigraph problem, called MultiGraphMatch;
\item An indexing data structure, called bit matrix, to efficiently select sets of candidate target edges to be matched with query edges;
\item A novel scoring scheme based on the cardinalities of compatibility domains to retrieve a heuristically promising order of processing query edges for the matching process;
\item A set of symmetry breaking conditions on query nodes and edges to filter out redundant matches during the search;
\item 
Software encompassing the above algorithms which will be available upon publication.
\end{itemize}

%\color{black}

To fully support queries on multigraphs, MultiGraphMatch uses the declarative SQL-like language CYPHER \cite{francis2018cypher} which allows: (i) the specification of logical conditions on both labels and properties; (ii) the formation of different kinds of outputs from the set of matches found.

We compare MultiGraphMatch with other systems that support queries on multigraphs, such as SuMGra \cite{ingalalli2016sumgra} as well as the graph databases Neo4J and Memgraph. Experiments show that MultiGraphMatch is generally faster than the compared systems on both synthetic and real networks and for both sparse and dense queries of different sizes.

The paper is organized as follows. Section \ref{related} describes related works for the subgraph matching problem in both simple graphs and multigraphs. Section \ref{definitions} defines preliminary concepts regarding multigraphs and the SMM problem. Section \ref{cypher} briefly describes the main features of the CYPHER query language. Section \ref{algorithm} explains each step of the MultiGraphMatch algorithm. Section \ref{experiments} shows experimental results performed on both synthetic and real networks. Finally, in Section \ref{conclusion}, we conclude the paper.

\section{Related Work}
\label{related}
Subgraph Matching (SM) is known to be an NP-complete problem \cite{cook1971complexity} and has attracted much research in the case of simple graphs. In the literature, four major classes of algorithms have been proposed for the SM problem based on the paradigm used to solve the matching problem: tree search algorithms, constraint propagation algorithms, indexing approaches and pre-processing methods.

In Tree Search (TS) algorithms the matching problem is defined in terms of a State Space Representation (SSR) and the algorithm explores a search space tree. Each state of the SSR corresponds to a partial solution. The search space is visited in a depth-first order and heuristics are used to prune parts of the search space tree to limit computational complexity. Backtracking is performed whenever the end of a branch is reached or pruning is done. Examples of TS algorithms include Ullmann’s algorithm \cite{ullmann1976algorithm}, VF2 \cite{cordella2004subgraph}, VF3 \cite{carletti2018challenging} and RI \cite{bonnici2013subgraph,bonnici2017variable,aparo2019simple}.

Constraint propagation algorithms model the SM problem as a constraint satisfaction problem, where query nodes denote the variables and target nodes represent values that can be assigned to such variables. Edges are translated into constraints that must be satisfied. These algorithms first compute, for each query node, a compatibility domain, which contains the set of target nodes that could be matched with that query nodes. Then, iteratively propagate the constraints to reduce such domains and filter candidate nodes for matching. Constraint propagation algorithms include nRF+ \cite{larrosa2002constraint}, Focus-Search \cite{ullmann2011bit}, ILF \cite{zampelli2010solving}, LAD \cite{solnon2010alldifferent} and Glasgow \cite{mccreesh2020glasgow}.

Indexing algorithms search for all graphs that match a given query from a database of targets by using an indexing-filtering-verification approach. In the indexing phase, data structures are constructed from characteristic substructures (e.g. paths, trees) of targets. In the filtering step, given a query graph $q$, target subgraphs are discarded if they have missing features contained in $q$.  Finally, a standard subgraph isomorphism algorithm is run to check which filtered target subgraphs fully match the query. Examples of indexing approaches are GraphQL \cite{he2008graphs}, SPath \cite{zhao2010graph}, GADDI \cite{zhang2009gaddi}, QuickSI \cite{shang2008taming} and Grapes \cite{giugno2013grapes}.

Pre-processing methods use auxiliary data structures built from the query and the target in order to select candidate nodes for the match and produce effective matching orders. Some of these algorithms also identify static and dynamic equivalences in the structure of the query and the target to eliminate redundant computations. Algorithms of this class include BoostISO \cite{ren2015exploiting}, TurboISO \cite{han2013turboiso}, CFL-Match \cite{bi2016efficient}, CECI \cite{bhattarai2019ceci}, DAF \cite{han2019efficient} and VEQ \cite{kim2022fast}.

The few comparison studies of algorithms proposed to solve the SM problem usually focus on the main existing tools \cite{aparo2019simple,sun2020inmemory,zeng2021deep}. Those studies show that indexing methods are more effective in pruning the search space, while pre-processing methods reduce the time of the matching phase, especially for large queries. Static processing order of query nodes (i.e. considering only information about the query) can also be effective in speeding up the search.

While the SM problem has been extensively studied, few solutions have been proposed for the SubMultigraph Matching (SMM) problem. The algorithmic paradigms proposed to solve the SM problem can apply to the SMM problem as well, but can be enhanced by using proper indexes to take advantage of the information associated with nodes and edges in order to speed up the matching process.

Solutions proposed to solve the SMM problem can be classified into two categories: in-memory algorithms and graph database systems. 

In-memory algorithms include SuMGra \cite{ingalalli2016sumgra}, Moorman's \cite{moorman2021subgraph}, FG$q_{t}$-Match \cite{sun2021subgraph} and Anwar's \cite{anwar2022subgraph}. SuMGra \cite{ingalalli2016sumgra} builds two indexes based on specific multigraph properties of the target, such as node and edge multiplicities, i.e. the number of node labels and the number of edges linking two nodes, respectively. The first index is the vertex signature, which captures information about the labels of edges incident on target nodes along with their multiplicities. The second index is the vertex neighborhood index, which contains information about the neighbors of a given node $u$, the edges connecting $u$ to its neighbors and the labels of each edge. Indexes are then used to select candidate nodes for the initial query node and the subsequent nodes during the search phase. Moorman's algorithm \cite{moorman2021subgraph} introduces a series of filtering methods to solve subgraph matching and similar problems on multigraphs. Filters are defined based on node labels, node-level statistics, topology, arc consistency and local neighborhood of nodes. Filters are then applied iteratively until convergence, to narrow down the search space. The authors also present an isomorphism counting approach that can be applied after filtering to count the number of query occurrences. FG$q_{t}$-Match \cite{sun2021subgraph} builds a matching-driven flow graph, called FG$q_{t}$, to perform subgraph matching on knowledge graphs. Nodes of FG$q_{t}$ are candidate target nodes for matching based on defined label constraints and each edge between a pair of candidates is redirected to an edge of FG$q_{t}$. A heuristic approach based on a multi-label weight matrix is followed to reduce the number of candidates. Matching is then performed by directly traversing FG$q_{t}$.

Anwar et al. \cite{anwar2022subgraph} propose a node embedding-based algorithm that utilizes two indexes computed on the target graph: the KD-tree, to represent node features, and the set-trie, to capture multi-edge features. These indexes are used to identify candidate sets for matching query nodes. To reduce the size of the node embeddings, the authors apply dimensionality reduction using Principal Component Analysis (PCA). Redundant solutions are avoided by enforcing symmetry-breaking conditions on the query graph. The authors show that their algorithm performs well on graphs with up to 1,000 nodes. In contrast, our focus is on graphs containing millions of nodes.

Graph database systems are databases in which data are modeled as directed and labeled multigraphs and data manipulation is done using graph-oriented operations, including subgraph matching \cite{timon2021overview}. Examples of graph database systems include Neo4J (\url{https://neo4j.com/}), Memgraph (\url{https://memgraph.com/}), ArangoDB (\url{https://www.arangodb.com}), OrientDB (\url{https://orientdb.org}), JanusGraph (\url{https://janusgraph.org}), TigerGraph \cite{deutsch2019tigergraph} and Titan (\url{https://titan.thinkaurelius.com}). Many systems rely upon storage design schemes of relational and NoSQL databases, such as key-value, document, tuple and object-oriented stores. However, traditional schemes are not suitable in densely connected data like graphs, where most relationships are many-to-many and querying the database requires multiple expensive join operations, impacting the performance. In contrast, native graph databases like TigerGraph, Memgraph and Neo4j are specifically built to maintain and process graphs efficiently. Labeled Property Graphs (LPG) and Resource Description Framework (RDF) schemes \cite{alaoui2019categorization,vilaca2010expressiveness} are commonly used for data modeling. Most databases use adjacency lists to represent graphs in the storage layer to make traversal operations faster. Data replication is ensured to enable data distribution and parallelization. Many graph databases also provide external interfaces, specific declarative languages to define queries (e.g. AQL, CYPHER), query optimizers and transaction engines.

\section{Preliminary Definitions}
\label{definitions}
In this section, we formally define a multigraph and the subgraph matching problem in multigraphs.

Given a set $X$, we denote by $\mathcal{P}(X)$ the power set of $X$.

\begin{definition}
A multigraph is a tuple $G=(V,E,\mathcal{L},\sigma,\mathcal{T},\pi,\mathcal{A},\mathcal{B})$, where:
\begin{itemize}
\item $V$ is the set of nodes, each of which has a unique identifier;
\item $E \subseteq N \times V \times V$ is the set of edges, \textcolor{black}{where $N$ is the set of unique edge identifiers}.  $N$ is necessary because two or more edges with different identifiers may be associated with the same pair of nodes;
\item $\mathcal{L}$ is the set of node labels;
\item $\sigma : V \rightarrow \mathcal{P}(\mathcal{L}) \setminus \{ \emptyset \}$ is the node label function, which associates one or more distinct labels from $\mathcal{L}$ to each node;
\item $\mathcal{T}$ is the set of edge labels, also called types;
\item $\pi : E \rightarrow \mathcal{T}$ is the edge type function, which associates one type from $\mathcal{T}$ to each edge;
\item $\mathcal{A} : V \rightarrow \mathcal{P}(P_V \times X_V) \setminus \{\emptyset\}$ is the node property function, which associates to each node a set of property-value pairs $\{(a_1, x_1),...,(a_l, x_l)\}$;
\item $\mathcal{B} : E \rightarrow \mathcal{P}(P_E \times Y_E) \setminus \{\emptyset\}$ is the edge property function, which associates to each edge a set of property-value pairs $\{(b_1, y_1),...,(b_m, y_m)\}$;
\end{itemize}
\end{definition}

Based on this definition, a multigraph is a graph with (potentially) multiple edges between two nodes, in which nodes and edges are annotated with labels and types, respectively, denoting classes they belong to.

In addition, both nodes and edges are associated with properties having specific values, which represent almost unique features of these objects. The definition of multigraph presented here is a generalization of the property graph \cite{angles2018property}, although the latter is not a multigraph.

We denote by $id(u)$ the identifier of node $u$. Similarly, we denote by $id(e)$ the identifier of edge $e$. Given an edge $e=(u,v)$, $u$ is the \textit{source} of $e$, $v$ is the \textit{destination} of $e$ and $u$ and $v$ are the \textit{endpoints} of $e$. If $(u,v) \in E$, we say that $u$ and $v$ are \textit{neighbors} and with $Neigh(u)$ we denote the set of all neighbors of node $u$. The \textit{out-degree} of a node $u$, $outDeg(u)$, is the number of edges having $u$ as source. Likewise, the \textit{in-degree} of $u$, $inDeg(u)$, is the number of edges having $u$ as destination. Given an edge type $t$, the \textit{$t$-dependent out-degree (in-degree)} of $u$, $t-outDeg(u)$ ($t-inDeg(u)$), is the number of edges of type $t$ having $u$ as source (destination). 

Given a node $u$, we denote as $\mathcal{A}(u)=\{(a_1, x_1),...,(a_l, x_l)\}$ the \textit{property set} of $u$. Likewise, given an edge $e$ of $G$, we denote by $\mathcal{B}(e)=\{(b_1, y_1),...,(b_m, y_m)\}$ the property set of $e$.

In what follows, we formally define the SubMultigraph Matching (SMM) problem.

\begin{definition}[SubMultigraph Matching (SMM)]
Given two multigraphs $Q=(V_Q,E_Q,\mathcal{L}_Q,\sigma_Q,$ $\mathcal{T}_Q,\pi_Q,\mathcal{A}_Q,\mathcal{B}_Q)$ and $T=(V_T,E_T,\mathcal{L}_T,\sigma_T,\mathcal{T}_T,\pi_T,$ $\mathcal{A}_T,\mathcal{B}_T)$, called query and target, respectively, the SubMultigraph Matching (SMM) problem consists in finding an injective function $f: V_Q \rightarrow V_T$, called node mapping, and an injective function $g: E_Q \rightarrow E_T$, called edge mapping, such that the following conditions hold:

\begin{enumerate}
\item \textcolor{black}{$\forall\, e=(x,u,v) \in E_Q$, $g(e)=(y,f(u),f(v)) \in E_T$;}
\item $\forall\, u \in V_Q$, $\sigma(u) \subseteq \sigma(f(u))$;
\item $\forall\, e \in E_Q$, $\pi(e) = \pi(g(e))$;
\item $\forall\, u \in V_Q$, $\forall\, (p,v) \in \mathcal{A}_Q(u)$ $(p,v) \in \mathcal{A}_T(f(u))$;
\item $\forall\, e \in E_Q$, $\forall\, (p,v) \in \mathcal{B}_Q(e)$ $(p,v) \in \mathcal{B}_T(g(e))$;
\end{enumerate}
\end{definition}

Condition 1 ensures that edge mapping $g$ is consistent with node mapping $f$, \textcolor{black}{i.e. neighbor nodes in the query are mapped to neighbor nodes in the target}. Condition 2 states that all the labels of a query node must be present among the labels of the corresponding mapped target node. Likewise, condition 3 states that mapped query and target edges must have the same type. Condition 4 implies that all property-value pairs of each query node be also present in the mapped target node. A similar constraint is expressed by condition 5 for the property set of mapped query and target edges.

Given an edge mapping $g$, an \textit{occurrence} of $Q$ in $T$ is a graph $O$ formed by edges $g(e_1), g(e_2), ..., g(e_k)$ and all nodes that are sources or destinations of at least one of these edges.

\begin{figure}
\centering
\includegraphics[width=12cm]{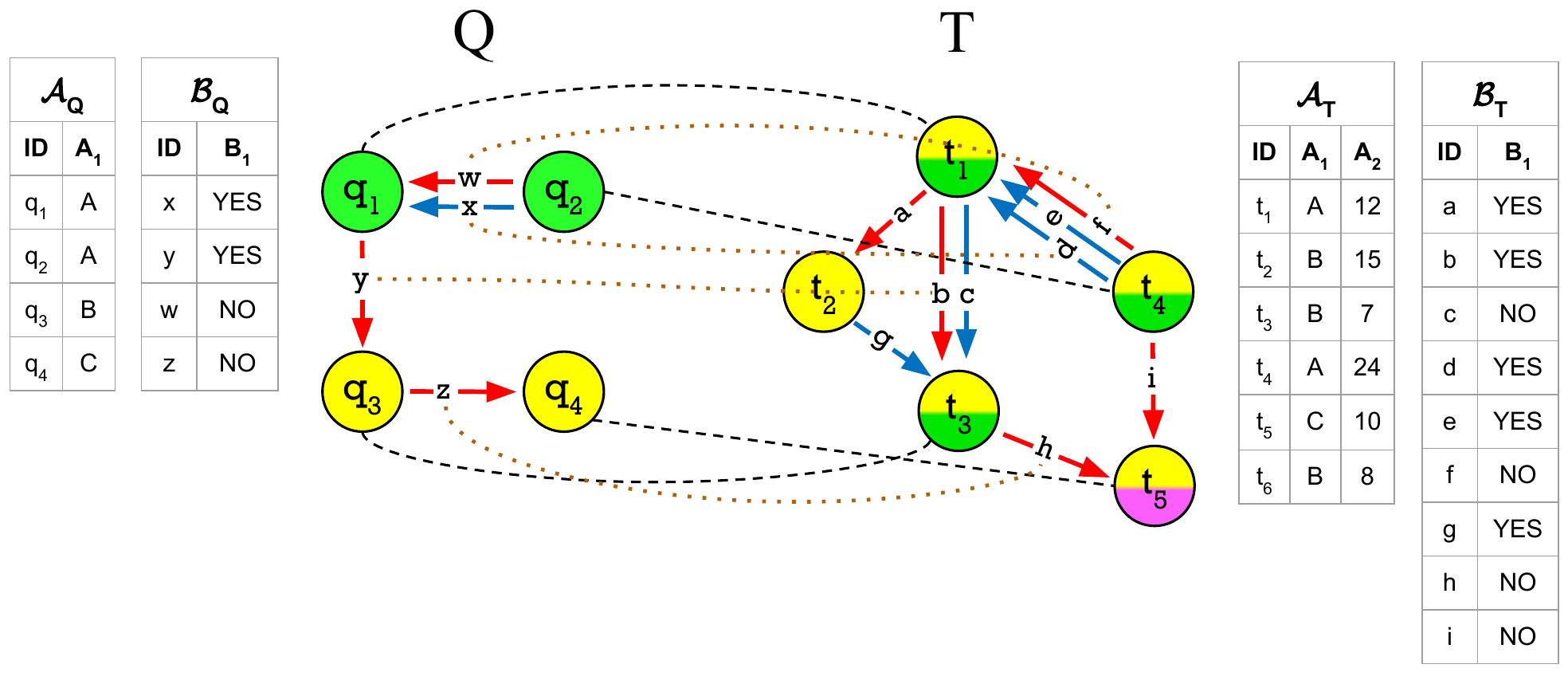}
  \caption{Toy example of SubMultigraph Matching (SMM) with a query $Q$ and a target $T$. Colors in nodes and edges represent different labels or types. Target nodes $t_1$, $t_3$, $t_4$ and $t_5$ have two labels. Black dashed lines denote a possible node mapping, with query nodes $q_1$, $q_2$, $q_3$ and $q_4$ mapped to target nodes $t_1$, $t_4$, $t_3$ and $t_5$, respectively. Orange dotted lines depict a possible edge mapping, with query edges $w$, $x$, $y$ and $z$ mapped to target edges $f$, $d$, $b$ and $h$, respectively.}
  \label{SIM_Example}
  \Description[SMM toy example]{Toy example of SubMultigraph Matching (SMM) with a query $Q$ and a target $T$}
\end{figure}

Fig. \ref{SIM_Example} shows an example of application of the SMM problem, where a possible node mapping is represented by black dashed lines and a possible edge mapping is depicted by orange dotted lines. The occurrence associated to these mappings is the subgraph of $T$ formed by nodes $t_1$, $t_3$, $t_4$ and $t_5$ and edges $d$, $f$, $b$ and $h$.

\section{CYPHER Query Language}
\label{cypher}
Queries in MultiGraphMatch are defined using the CYPHER query language \cite{francis2018cypher}. CYPHER was originally integrated within the graph database system Neo4J\footnote{https://neo4j.com/}, but in 2015 has become an open language thanks to the openCypher project\footnote{https://opencypher.org/}. It extends the capabilities of the Structured Query Language (SQL), designed for managing and querying data stored in relational databases. 

CYPHER language can be used to define query graphs in a multigraph that match specific topological constraints and optionally semantic conditions on node labels, edge types and properties. MultiGraphMatch implements a subset of the CYPHER query language related to the SMM problem. A complete specification of the CYPHER language is available at \url{https://opencypher.org/resources}. 

Here, we briefly summarize the main features of CYPHER used in this paper.

A complete specification of CYPHER's syntactic constructs supported by MultiGraphMatch is available at \href{https://github.com/Anto188bas/MultiGraphMatch.git}{https://github.com/Anto188bas/MultiGraphMatch.git}. The example queries in the CYPHER language presented here are taken from the \textsc{imdb} bipartite multigraph used in our experiments (see Section \ref{experiments}). In \textsc{imdb} the two classes of nodes are people involved in movie industry (e.g. actors, directors, composers) and movies. Directed edges connect people to movies. 

The heart of the CYPHER query language is represented by the following three clauses: \texttt{MATCH}, \texttt{WHERE}, and \texttt{RETURN}. The \texttt{MATCH} clause is used to define the query's topology. The \texttt{WHERE} clause is used to optionally filter occurrences of a query based on conditions imposed on labels, types or properties. The \texttt{RETURN} clause specifies the type of output we want to extract from the set of query's occurrences found.

In the \texttt{MATCH} clause the query's topology is defined as a list of edges. Nodes forming an edge must be enclosed in round (parenthetic) brackets, while edges have to be enclosed within square brackets. Nodes and edges can be named and referenced multiple times within the \texttt{MATCH} clause (e.g. if they are involved in more than one edge or path). For query nodes (or edges) that are defined for the first time in the \texttt{MATCH} clause, we can also specify their labels (or types) by using semicolons. Node or edge properties can be indicated as well, by means of the notation \texttt{\{p1:v1, p2:v2, ..., pk:vk\}}, where \texttt{p1, p2, ..., pk} and \texttt{v1, v2, ..., vk} are the property names and their corresponding values. Edges are linked to nodes by using one of the following strings: "-", "<-" or "->". The latter two strings are used to create directed or bi-directional links.

For instance, the following \texttt{MATCH} clause:
\\ \\
\texttt{MATCH (a:director)-[:DIRECTED \{year:2007\}]->(b:drama), \\ (c {nationality:USA})-[:DIRECTED]->(b)}
\\ \\
specifies a query in which we are looking for drama movies, referenced as $b$ in the query, that were co-directed in 2007 by two different directors, named $a$ and $c$, one of which (node $c$) is American.

The \texttt{WHERE} clause is optional and is formed by a list of conditions on labels, types and properties. Conditions are expressed using relational operators (e.g. $<, \leq, >, \geq, !=$) or special string operators, such as \texttt{STARTS WITH}, \texttt{ENDS WITH} and \texttt{CONTAINS}. Logical operators (e.g. \texttt{NOT}, \texttt{AND}, \texttt{OR}) can be used to combine conditions. The following is an example of a CYPHER query with a \texttt{WHERE} clause:
\\ \\
\texttt{MATCH (a:producer)-[r:PRODUCED]->(b:adventure) \\ WHERE a.surname STARTS WITH 'Sp' AND r.year >= 2006}
\\ \\
In this case, we are looking for producers who produced adventure movies in 2006 or later and whose last name starts with 'Sp' (e.g.  Steven Spielberg). Note that all nodes or edges for which we want to impose conditions must be previously referenced in the \texttt{MATCH} clause.

The \texttt{RETURN} clause is always mandatory. The information returned by the clause can be either the number of occurrences found (using the \texttt{count()} function) or a table containing the values of labels (using the function \texttt{labels()}), types (using the function \texttt{type()}) or properties of one or more nodes or edges involved in the occurrences found, provided that the latter have been previously referenced in the \texttt{MATCH} clause. A list of nodes and edges forming each occurrence can be obtained using functions \texttt{nodes()} and \texttt{relationships()}. The set of a query's occurrences to be returned can be limited to the first $k$ occurrences using \texttt{LIMIT k}. For instance, in the query:
\\ \\
\texttt{MATCH (a:writer)-[r:WROTE]->(b:thriller) \\ RETURN a.surname, b.name, r.year LIMIT 10}
\\ \\
we look for writers of thriller movies and output the writer's surname, the movie's name and the year of writing. In addition, we are limiting our search to the first 10 occurrences found.

\section{Description of MultiGraphMatch}
\label{algorithm}
In this section, we provide the details of MultiGraphMatch algorithm, which has been implemented in the Java language. MultiGraphMatch takes as input a target and a query written in CYPHER language. 

Before matching the query, MultiGraphMatch creates different data structures to index the whole target network into main memory. Then, the algorithm performs the following steps: 
\begin{enumerate}
\item[a)] Computation of symmetry breaking conditions for query nodes and edges;
\item[b)] Computation of compatibility domains;
\item[c)] Ordering of processing query edges for the matching;
\item[d)] Matching query with target.
\end{enumerate}

In the next subsections, we detail each of these steps. For ease of explanation, we will consider queries with no \texttt{WHERE} clause. In the last subsection, we will illustrate how to adapt MultiGraphMatch to handle logical conditions expressed by the \texttt{WHERE} clause. For the description of the algorithm we refer to a query $Q=(V_Q,E_Q,\mathcal{L}_Q,\sigma_Q,\mathcal{T}_Q,\pi_Q,\mathcal{A}_Q,\mathcal{B}_Q)$ with $k$ nodes and a target \\ $T=(V_T,E_T,\mathcal{L}_T,\sigma_T,\mathcal{T}_T,\pi_T,\mathcal{A}_T,\mathcal{B}_T)$ with $n$ nodes. In order to explain each step, we will also refer to the toy example query $Q$ and target $T$ of Fig. \ref{SIM_Example}.

\subsection{Indexing of target network}
\label{IndexingSec}

As a preprocessing step, the target is read into main memory and several data structures are created to efficiently index it and speed up the next phases of the algorithm, in particular the computation of compatibility domains and the matching process.

The four data structures created are:
\begin{enumerate}
\item[a)] Node label graph;
\item[b)] Edge types map;
\item[c)] Edge properties table;
\item[d)] Bit signature matrix.
\end{enumerate}

\begin{figure}
\centering
\includegraphics[width=11cm]{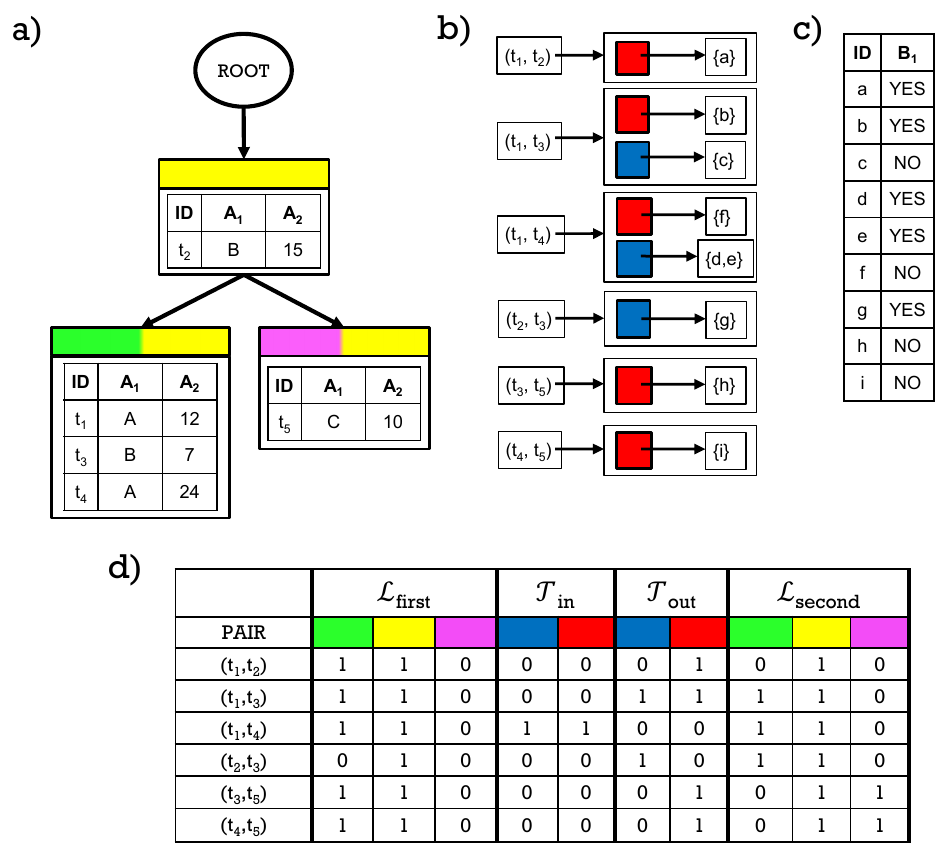}
  \caption{Indexing data structures associated to the target $T$ of Fig. \ref{SIM_Example}: a) Node label graph, b) Edge types map, c) Edge properties table, d) Bit signature matrix.}
  \label{indexing}
  \Description[Indexing data structures]{Indexing data structures associated to the target $T$ of Fig. \ref{SIM_Example}}
\end{figure}

An example of computation of these indexing data structures is shown in Fig. \ref{indexing} for the target $T$ of Fig. \ref{SIM_Example}.

The node label graph (Fig. \ref{indexing}a) is a graph in which each vertex\footnote{To avoid confusion, we use here the term "vertex" to denote a node of the node label graph and the term "node" to indicate a query node or a target node.} is decorated with a specific set of labels. There is one vertex for each subset of labels that are present in the target graph. In the example of Fig. \ref{SIM_Example} the target nodes have three distinct sets of labels ("yellow", "green and yellow" and "red and yellow"), so the corresponding node label graph (Fig. \ref{indexing}a) has three different vertices. A vertex in the node label graph decorated with the set of labels $L$ contains a table with the ids of nodes having labels $L$ and their associated properties. For example $t_1$, $t_3$, and $t_4$ all have green and yellow labels, but differ in their property values. A directed edge links vertex $V_1$ to vertex $V_2$ iff the set of labels of nodes in $V_1$ is a subset of the set of labels of nodes in $V_2$. For example, target node $t_2$ has only a yellow label while $t_1$, $t_3$, and $t_4$ have yellow and green labels. In the worst case, the node label graph has a number of vertices $K$ equal to the minimum of the size of the power set of all labels and the number of target nodes.

The edge types map (Fig. \ref{indexing}b) is a two-level hash map. The outer level associates an ordered pair of nodes $(t_i,t_j)$ to an inner local hash map where each key is an edge type $X$ and the corresponding value is the list of all edges linking $t_i$ to $t_j$ and having type $X$.

The edge properties table (Fig. \ref{indexing}c) is a table that lists properties for each target edge.

The bit signature matrix (Fig. \ref{indexing}d) is a matrix of bits where each row $R$ contains, for a given pair of connected nodes $(t_i,t_j)$ in the target (with $id(t_i)<id(t_j)$), a string of bits, called the \textit{signature} of row $R$ and denoted as $Sign(R)$. The signature is a concatenation of four substrings of bits indicating the labels of $t_i$ and $t_j$ and the types of all edges connecting the two nodes in both directions.

Formally, suppose that there are $h$ node labels and $p$ edge labels in the target graph and that both target labels and types are arbitrarily ordered. The bit signature of the row $R$ associated with the pair $(t_i,t_j)$ is a string of bits $\mathcal{L}_{first}\cdot\mathcal{T}_{in}\cdot\mathcal{T}_{out}\cdot\mathcal{L}_{second}$, where $\cdot$ is the string concatenation operator and:

\begin{itemize}
\item $\mathcal{L}_{first}$ is a substring of bits $b_{11}b_{12} \cdots b_{1h}$, where $b_{1k}=1$ iff node $t_i$ contains the $k$-th label in the ordered set of node labels and $b_{1k}=0$ otherwise;
\item $\mathcal{T}_{in}$ is a substring of bits $b_{21}b_{22} \cdots b_{2p}$, where $b_{2k}=1$ iff there is at least one directed edge $(t_j,t_i)$ having the $k$-th type in the ordered set of edge types and $b_{2k}=0$ otherwise;
\item $\mathcal{T}_{out}$ is a substring of bits $b_{31}b_{32} \cdots b_{3p}$, where $b_{3k}=1$ iff there is at least one directed edge $(t_i,t_j)$ having the $k$-th type in the ordered set of edge types and $b_{3k}=0$ otherwise;
\item $\mathcal{L}_{second}$ is a substring of bits $b_{41}b_{42} \cdots b_{4h}$, where $b_{4k}=1$ iff node $t_j$ contains the $k$-th label in the ordered set of node labels and $b_{4k}=0$ otherwise;
\end{itemize}

Intuitively, the bit signature matrix is a sort of one-hot encoding scheme of the node labels and edge types associated with each pair of connected nodes in the target. Note that the bit signature matrix records only the presence or absence of edges with a specific type and direction linking two nodes and not the number of such edges. However, the purpose of the bit signature matrix is to quickly filter target candidate nodes for matching during the computation of compatibility domains in the next steps, so this data structure represents a compromise between memory use and filtering capabilities. Whereas some state-of-the-art systems use more complex indexes to reduce candidates, particularly in multigraphs, our experiments show that these simpler indexes result in better overall performance.

\subsection{Computation of symmetry breaking conditions}
\label{breakingCondSec}

Several mappings in the SMM problem may in fact correspond to the same occurrence in the target.

\begin{figure}
\centering
\includegraphics[width=11cm]{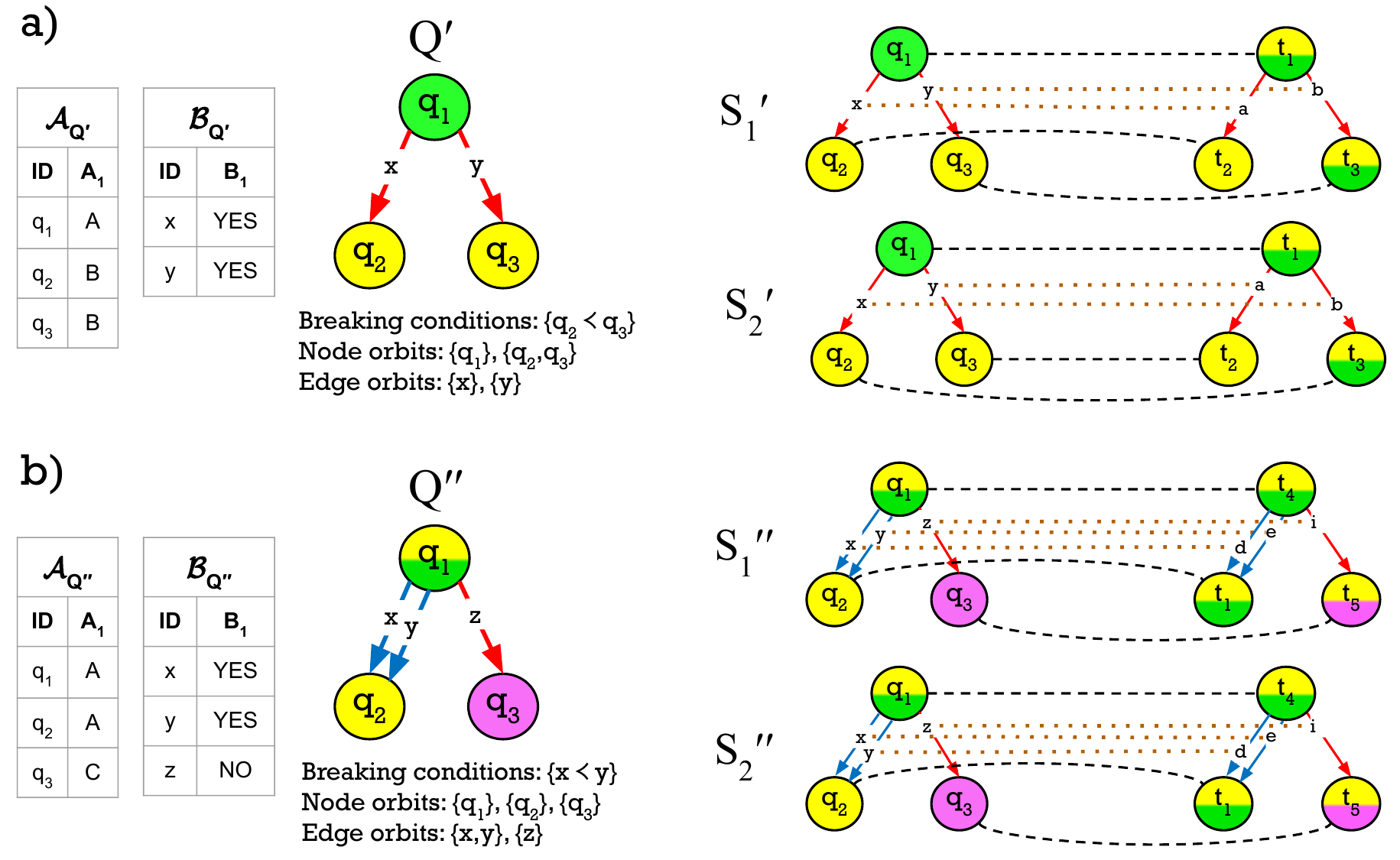}
  \caption{Example of symmetry breaking condition on nodes and edges and application of conditions on matching queries $Q'$ and $Q''$ with the target multigraph $T$ of Fig. \ref{SIM_Example}. a) In query $Q'$ nodes $q_2$ and $q_3$ are in the same orbit, thus yielding the breaking condition $q_2 \prec q_3$. Assuming that $t_2<t_3$ in $T$, after the application of such condition on $T$, solution $S'_2$ is discarded. b) In query $Q''$ edges $x$ and $y$ are in the same orbit, thus yielding the breaking condition $x \prec y$. Assuming that $d<e$ in $T$, after the application of such condition on $T$, solution $S''_2$ is discarded.}
  \label{breakingCond}
  \Description[Breaking conditions example]{Example of symmetry breaking condition on nodes and edges and application of conditions on matching queries $Q'$ and $Q''$ with the target multigraph $T$ of Fig. \ref{SIM_Example}}
\end{figure}

Fig. \ref{breakingCond} illustrates two toy examples with the same target $T$ of Fig. \ref{SIM_Example} and two queries $Q'$ and $Q''$. In the query $Q'$ (Fig. \ref{breakingCond}a), nodes $q_2$ and $q_3$ have the same sets of labels and are both connected to $q_1$ with an edge having the same direction and type. They can be considered equivalent and can be mapped to either target node $t_2$ or target node $t_3$. This results in two distinct mappings that correspond to the same occurrence in $T$. Therefore, one of them should be excluded from the set of solutions. Likewise, in the query $Q''$ (Fig. \ref{breakingCond}b), edges $x$ and $y$ have the same type, the same direction and connect the same pair of nodes. So, they can be indifferently mapped to target edges $d$ and $e$. The resulting mappings correspond to the same occurrence and one of them can be safely discarded.

To avoid redundancies, MultiGraphMatch computes \textit{symmetry breaking conditions} on query nodes and edges.

In the literature, breaking conditions have been mainly used in motif search \cite{grochow2007network,ribeiro2014gtries} and subgraph matching \cite{demeyer2013index,houbraken2014index,han2019efficient,kim2022fast,yang2023structural} on simple graphs. Our novel contribution is to extend the concept of breaking conditions to the SMM problem.

As a short review, symmetry breaking conditions are related to the concepts of automorphisms and orbits of a multigraph. 

An automorphism is a permutation of nodes and edges of a multigraph that preserves the structure of the graph, the labels and the properties of its nodes and the types and the properties of its edges. Automorphism is an equivalence relation that induces a partition of the sets of multigraph's nodes and edges, respectively, into equivalence classes, called \textit{orbits}. In other words, nodes (or edges) in the same orbit can be permuted without affecting the structure or the semantics of the multigraph. Symmetry breaking conditions are inequalities of the form $a \prec b$ that impose a relative order between two query nodes (or two query edges) belonging to the same orbit.

In the query graph $Q'$ of Fig. \ref{breakingCond}a) nodes $q_2$ and $q_3$ are in the same orbit, therefore we can define a symmetry breaking condition $q_2 \prec q_3$. Similarly, in Fig. \ref{breakingCond}b) edges $x$ and $y$ belong to the same orbit, so we can define the breaking condition $x \prec y$.

The computation of symmetry breaking conditions is done by extending the method presented in \cite{ribeiro2014gtries} for simple unlabeled graphs. First, query automorphisms and orbits are computed using the NAUTY algorithm \cite{mckay2014practical}. Then, breaking conditions are extracted from equivalence classes with at least two nodes or edges, always starting from the element having the smallest id.

Symmetry breaking conditions are applied to target nodes and edges during the matching process to discard redundant occurrences. Specifically, let $q$ and $q'$ be two query nodes, $t$ and $t'$ be two target nodes, and $f$ be a node mapping function which is a solution to the SMM problem, with $f(q)=t$ and $f(q')=t'$. If $q \prec q'$, then $q$ and $q'$ are in the same orbit, so either can be mapped to $t$ and the other to $t'$. As a result, there must exist another node mapping function $f'$ that yields a second solution to the SMM problem, with $f'(q)=t'$ and $f'(q')=t$. The two solutions produce the same occurrence and in one of them $id(t)<id(t')$, while in the other one $id(t)>id(t')$. To discard one of the two redundant solutions, we simply require that $id(t)<id(t')$. Similar reasoning applies to edge breaking conditions.

In Fig. \ref{breakingCond}a, assuming $t_2<t_3$, the application of the breaking condition $q_2 \prec q_3$ results in including solution $S'_1$ and discarding solution $S'_2$. In Fig. \ref{breakingCond}b, assuming $d<e$, the application of the breaking condition $x \prec y$ results leads to include solution $S''_1$ and discard solution $S''_2$.

\subsection{Computation of compatibility domains}
\label{CompDomainsSec}

Before matching the query with the target, MultiGraphMatch builds a set of matchable target edges based on compatibility domains for all pairs of connected query nodes. Only matchable edges will be considered as candidates for a match during the search.

Given the pair of connected query nodes $(q_i,q_j)$, with $id(q_i)<id(q_j)$, the \textit{compatibility domain} of $(q_i,q_j)$, $Dom(q_i,q_j)$, is a set of pairs of connected target nodes $(t_i,t_j)$ which are selected according to a set of conditions involving: i) the direction and the type of all the edges connecting $q_i$ to $q_j$; ii) the labels of $q_i$ and $q_j$; iii) the type-dependent in-degrees and out-degrees of $q_i$ and $q_j$. If all conditions are satisfied, $(t_i,t_j)$ is said to be \textit{compatible} to $(q_i,q_j)$ and the pair $(t_i,t_j)$ is added to $Dom(q_i,q_j)$. Node and edge properties are not used to build compatibility domains.

Conditions based on the direction and types of query edges and the labels of query nodes are checked by using the bit signature matrix data structure introduced in Subsection \ref{IndexingSec}. Specifically, the algorithm computes the bit signature matrix of query $Q$, $\mathcal{BS}(Q)$, and then matches the rows of $\mathcal{BS(Q)}$ with the rows of the bit signature matrix of the target $T$, $\mathcal{BS}(T)$. 

\begin{figure}
\centering
\includegraphics[width=11cm]{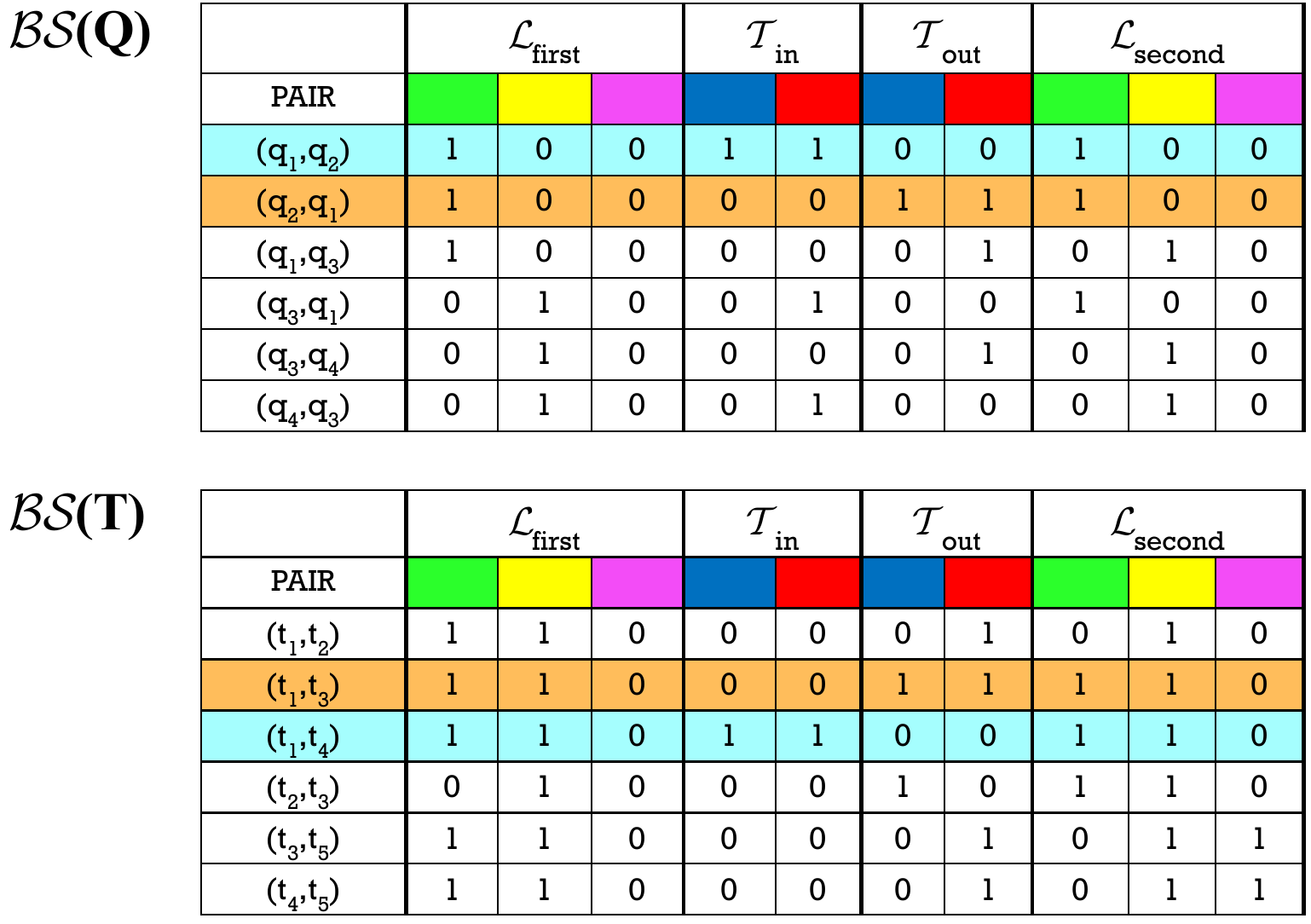}
\caption{Example of computation of compatibility
domains for the query $Q$ and target $T$ of Fig. \ref{SIM_Example}. $\mathcal{BS}(Q)$ and $\mathcal{BS}(T)$ are the bit signature matrices of $Q$ and $T$, respectively. The row of pair $(t_1,t_4)$ contains all the features of the row associated to $(q_1,q_2)$, so $(t_1,t_4)$ matches $(q_1,q_2)$. Target pair $(t_1,t_3)$ does not match $(q_1,q_2)$ but matches $(q_2,q_1)$. This is equivalent to say that $(t_3,t_1)$ matches $(q_1,q_2)$. In addition, target nodes of both pairs $(t_1,t_4)$ and $(t_3,t_1)$ have at least the same number of type-dependent in-degrees and out-degrees of corresponding query nodes $q_1$ and $q_2$. No more pairs of target nodes match either $(q_1,q_2)$ or $(q_2,q_1)$, because they miss a pair of blue and red edges in either ingoing or outgoing direction. In the case that $id(q_1)<id(q_2)$, we can build the compatibility domain of $(q_1,q_2)$, given by $Dom(q_1,q_2)=\{(t_1,t_4),(t_3,t_1)\}$.}
\label{domains}
\Description[Computation of compatibility domains]{Example of computation of compatibility
domains for the query $Q$ and target $T$ of Fig. \ref{SIM_Example}}
\end{figure}

Each row of $\mathcal{BS}(Q)$ contains the bit signature of a specific pair of connected query nodes. The structure of $\mathcal{BS}(Q)$ is the same as the bit signature matrix of the target (see Subsection \ref{IndexingSec}), with node labels and edge types following the same order chosen to build $\mathcal{BS}(T)$. The only difference is that now for the query graph, the bit signature matrix includes the edges when considering $q_i$ as source and $q_j$ as destination as well as vice versa. This is done to speed up the computation of domains and to ensure that all compatible pairs are discovered. Fig. \ref{domains} shows the bit signature matrices $\mathcal{BS}(Q)$ and $\mathcal{BS}(T)$ for the query $Q$ and target $T$ of Fig. \ref{SIM_Example}, respectively.

In order to find all pairs of connected target nodes that match the labels of $q_i$ and $q_j$ and the types and directions of all edges that link $q_i$ and $q_j$, MultiGraphMatch scans the bit signature matrix of target $\mathcal{BS}(T)$ to search for all rows that contain all the entries of the rows of either pair $(q_i,q_j)$ or pair $(q_j,q_i)$. Changing the order of target and query nodes in their respective pairs corresponds to simply swapping node labels and the direction of the edges connecting them both in the query and in the target, preserving their match. Therefore, if the row of $(t_i,t_j)$ contains all the entries of the row of $(q_j,q_i)$, this is equivalent to saying that $(t_j,t_i)$ matches $(q_i,q_j)$.

Mathematically, checking if a row $R$ contains all the entries in the row $R'$ corresponds to verify if $Sign(R) \land Sign(R') = Sign(R)$, where $\land$ is the bitwise AND operator.

Bit matrices record only the presence or absence of edges of a specific type. To also take into account the multiplicity of a type in the edges connecting two nodes, if the row of a target pair $(t_i,t_j)$ matches either the row of $(q_i,q_j)$ or the row of $(q_j,q_i)$, type-dependent in-degrees and out-degrees of the corresponding query and target nodes are compared. In particular, suppose that $q_i < q_j$ and the row of $(t_i,t_j)$ matches the row of $(q_i,q_j)$. Then, $(t_i,t_j)$ is added to $Dom(q_i,q_j)$ iff $\forall t \in \mathcal{T}$:

\begin{enumerate}
\item $t-outDeg(q_i) \leq t-outDeg(t_i)$;
\item $t-inDeg(q_i) \leq t-inDeg(t_i)$;
\item $t-outDeg(q_j) \leq t-outDeg(t_j)$;
\item $t-inDeg(q_j) \leq t-inDeg(t_j)$;
\end{enumerate}

In Fig. \ref{domains} the row corresponding to the pair $(t_1,t_4)$ contains all the entries of the row of $(q_1,q_2)$, so the two rows match. The row of $(t_1,t_3)$ would also match the row of $(q_1,q_2)$ if $t_1$ and $t_3$ were swapped. Therefore, the row of $(t_1,t_3)$ matches the row of $(q_2,q_1)$. Since both $t_1$ and $q_1$ have an outgoing red edge and $t_4$ have at least one red outgoing edge and one blue outgoing edge as does query node $q_2$, we can conclude that $(t_1,t_3)$ is compatible with $(q_1,q_2)$. By comparing the type-dependent in-degrees and out-degrees of the nodes $q_2$ and $t_1$ and the nodes $q_1$ and $t_3$, we can state that $(t_3,t_1)$ is compatible with $(q_1,q_2)$ as well. There are no other pairs of target nodes that either match $(q_1,q_2)$ or $(q_2,q_1)$. Provided that $id(q_1)<id(q_2)$, we can build a compatibility domain for the pair $(q_1,q_2)$, given by $Dom(q_1,q_2)=\{(t_1,t_4),(t_3,t_1)\}$.

\subsection{Determining the processing order of query edges}
\label{ordering}

To establish in which order query edges have to be processed during the matching, MultiGraphMatch exploits information about compatibility domains. As shown in \cite{bonnici2017variable},the processing order of query nodes and edges can have a large impact on the running time of subgraph matching algorithms. State-of-the-art algorithms typically begin with query nodes or edges that have the fewest candidates. However, in many cases, starting with a query edge involving high-degree nodes - even though these nodes do not have the smallest candidate domains - can lead to more effective and earlier pruning of the search space during matching. Our idea is to start by matching the densest (highest degree) parts of the query graph, while also taking into account the cardinalities of compatibility domains and the number of matching constraint based on the pairs that are already in the ordering. 

Starting from an empty {\em sequential ordering} $\mathcal{O}$, MultiGraphMatch iteratively chooses a new pair of connected query nodes $(q_i,q_j)$ (with $id(q_i)<id(q_j)$) and adds all edges linking $q_i$ and $q_j$ to $\mathcal{O}$ in arbitrary order. The process ends when all query edges have been added to $\mathcal{O}$.

At each step of the ordering, the pair of query nodes is chosen according to a priority score. Given a pair $(q_i,q_j)$ with $id(q_i)<id(q_j)$ the priority of $(q_i,q_j)$ is a pair $(CF,Sc)$, where:
\begin{itemize}
\item $CF$ is the constraint factor, defined as the number of nodes in $(q_i,q_j)$ that belong to pairs already present in the current ordering $\mathcal{O}$;
\item $Sc$ is a score which depends on the the density of the neighborhood of $(q_i,q_j)$ and the size of its compatibility domain.
\end{itemize}

$CF$ can take values 0, 1 or 2. In the case of $CF=1$, we call the node that does not yet belong to any pair already in the ordering {\em free node}. The score $Sc$ is defined differently for different values of  $CF$: 
\begin{equation}
Sc(q_i,q_j) = \begin{cases}
\frac{totDeg(q_i) \times totDeg(q_j) \times Jacc(q_i,q_j)}{|Dom(q_i,q_j)|} &\text{if $CF(q_i,q_j)=0$}\\
\frac{totDeg(freenode(q_i,q_j)) \times Jacc(q_i,q_j)}{|Dom(q_i,q_j)|} &\text{if $CF(q_i,q_j)=1$}\\
\frac{1}{|Dom(q_i,q_j)|} &\text{if $CF(q_i,q_j)=2$}\\
\end{cases}
\end{equation}

where $totDeg(q_i)$ is the total degree (i.e. out-degree + in-degree) of node $q_i$, $freenode(q_i,q_j)$ is the free node of pair $(q_i,q_j)$ and $Jacc(q_i,q_j)$ is the neighbor Jaccard similarity between $q_i$ and $q_j$, i.e. the fraction of neighbors in common between $q_i$ and $q_j$:
\begin{equation}
Jacc(q_i,q_j) = \frac{|Neigh(q_i) \cap Neigh(q_j)|}{|Neigh(q_i) \cup Neigh(q_j)|}
\end{equation}

Note that, except for the case $CF=2$, $Sc$ combines the domain size and the neighborhood's local density of a pair, so the score expression reflects a trade-off between the two measures. For example, it may be better to start from edges with very small domains even though they are not in a dense core region of the query.

To choose the next pair to be added to the current ordering, MultiGraphMatch first computes the priority of all pairs that are not yet in the current ordering $\mathcal{O}$. Next, candidate pairs are lexicographically compared: a pair $(u,v)$ has higher priority than pair $(u',v')$ iff either $CF(u,v)>CF(u',v')$ or $CF(u,v)=CF(u',v')$ and $Sc(u,v)>Sc(u',v')$. The pair with highest priority is added to the current ordering $\mathcal{O}$ and possible ties are solved arbitrarily. 

Thus, the algorithm gives higher priority to pairs having $CF=2$. This is done to extend a partial match from edges that have the maximum number of constraints as soon as possible. In fact, if both endpoints of a query edge $e$ have already been matched, there are likely very few candidate target edges to scan while matching $e$.

\subsection{Matching process}
\label{MatchingSec}

Following the previously defined ordering of query edges, MultiGraphMatch performs matching to find occurrences of the query within the target. The matching process is outlined in Algorithm \ref{SubgraphMatching_Algo}.

\IncMargin{1em}
\begin{algorithm}
\small
\SetKwInput{kwInit}{Init}
\KwIn{$Q=(V_Q,E_Q,\mathcal{L}_Q,\sigma_Q,\mathcal{T}_Q,\pi_Q,\mathcal{A}_Q,\mathcal{B}_Q)$: query, $T=(V_T,E_T,\mathcal{L}_T,\sigma_T,\mathcal{T}_T,\pi_T,\mathcal{A}_T,\mathcal{B}_T)$: target, $\mathcal{BC}_N$: set of breaking conditions on nodes, $\mathcal{BC}_E$: set of breaking conditions on edges, $Dom$: compatibility domains, $\mathcal{O}$: ordering of query edges}
\KwOut{$Occs$: list of occurrences}
\kwInit{$f(u) := undefined$ for all $u \in V_Q$\,\,\,\,\,\,//Current node mapping}
\kwInit{$g(e) := undefined$ for all $e \in E_Q$\,\,\,\,\,\,//Current edge mapping}
\kwInit{$Cand(e) := undefined$ for all $e \in E_Q$\,\,\,\,\,\,//Lists of candidate target edges}
\kwInit{$CandIndex(e) := 1$ for all $e \in E_Q$\,\,\,\,\,\,//Iterators for candidate lists}
$e_Q=(q,q')$ := $\mathcal{O}[1]$\;
$Cand(e_Q) :=$ \textsc{FindCandidates}($Q,e_Q,T,\mathcal{BC}_N,\mathcal{BC}_E,Dom,f,g$)\;
$i := 1$\;
\While{$i>0$}{
    \uIf{$CandIndex(e_Q)>|Cand(e_Q)|$}{
        RestoreInfo($f,g$)\;
        $i := i-1$\;
        $e := \mathcal{O}[i]$\;
    }
    \uElse{
        $e_T=(t,t') := Cand(e_Q)[candIndex(e_Q)]$\;
        \uIf{$\forall$ $e_Q \in E_Q: g(e_Q) \neq e_T$}{
            \uIf{$f(q) = undefined$}{
                $f(q) := t$\;
            }
            \uIf{$f(q') = undefined$}{
                $f(q') := t'$\;
            }
            $g(e_Q) := e_T$\;
            \uIf{$i=|E_Q|$}{
                $Occ :=$ BuildOccurrence($f,g$)\;
                $Occs := Occs \cup \{Occ\}$\;
                $CandIndex(e) := CandIndex(e)+1$\;
            }
            \uElse{
                $i := i+1$\;
                $e_Q=(q,q')$ := $\mathcal{O}[i]$\;
                $Cand(e_Q) :=$ \textsc{FindCandidates}($Q,e_Q,T,\mathcal{BC}_N,\mathcal{BC}_E,Dom,f,g$)\;
                $CandIndex(e_Q) := 1$\;
            }
        }
        \uElse{
           $CandIndex(e_Q) := CandIndex(e_Q)+1$\;
        }
    }
}
\KwRet{$Occs$}
\caption{\textsc{SubgraphMatching($Q,T,\mathcal{BC}_N,\mathcal{BC}_E,Dom,\mathcal{O}$)}}
\label{SubgraphMatching_Algo}
\end{algorithm}

Matching is done by building a node mapping function $f: V_Q \rightarrow V_T$ and an edge mapping function $g: E_Q \rightarrow E_T$. The list of candidates to scan during the search is stored in variable $Cand$.

The matching process starts with the first edge $e_Q=(q,q')$ in the ordering (line 1). An initial list of candidates $Cand(e)$ is calculated (line 2) and the list is scanned starting from the first element. A given candidate $e_T=(t,t')$ can be matched with $e_Q$ iff $e_T$ has not yet been mapped (line 11). Whenever a new match is found, the mapping functions $f$ and $g$ are updated (lines 12-16). If $e_Q$ was the last query edge to match, the occurrence $Occ$ resulting from mappings $f$ and $g$ is built and added to the list of matches found (lines 18-19). The search then continues with the next candidate (line 20). If $e_Q$ was not the last query edge to match, we consider the next query edge to match (lines 22-23) and find the set of candidates for such edge (line 24). If candidate $e_T$ does not match $e_Q$, the algorithm simply skips to the next candidate (line 27). When all candidates for $e_Q$ have been examined (line 5), MultiGraphMatch performs backtracking, i.e. restores mappings to the previous values (line 6) and goes back to the previous query edge (lines 7-8). Backtracking implies removing the mapping for the last matched query edge $e_q$ and optionally the mapping for one or both nodes of $e_q$. To guarantee that every time we do backtracking the search continues from last examined candidate for the previous query edge, MultiGraphMatch uses a set of list iterators $CandIndex$, one for each query edge. $CandIndex(e_Q)$ contains the position of the last candidate examined in $Cand(e_Q)$. Every time we pass to the next candidate, the corresponding iterator is incremented (lines 20 and 27). The search ends when no more candidates are available for the first query edge. At the end of the process, MultiGraphMatch returns the list of all occurrences found (line 28).

The procedure used to find the set of candidates $Cand(e_Q)$ for a query edge $e_Q=(q,q')$ is detailed in Algorithm \ref{FindCandidates_Algo}.

\IncMargin{1em}
\begin{algorithm}
\small
\KwIn{$Q=(V_Q,E_Q,\mathcal{L}_Q,\sigma_Q,\mathcal{T}_Q,\pi_Q,\mathcal{A}_Q,\mathcal{B}_Q)$: query, $e_Q=(q,q')$: query edge, $T=(V_T,E_T,\mathcal{L}_T,\sigma_T,\mathcal{T}_T,\pi_T,\mathcal{A}_T,\mathcal{B}_T)$: target, $\mathcal{BC}_N$: set of breaking conditions on nodes, $\mathcal{BC}_E$: set of breaking conditions on edges, $Dom$: compatibility domains, $f$: current node mapping, $g$: current edge mapping}
\KwOut{$Cand$: list of candidates}
\uIf{$f(q) = undefined \land f(q') = undefined$}{
    \ForEach{$e_T=(t,t') \in Dom(q,q')$}{
        \uIf{$\pi_T(e_T)=\pi_Q(e_Q) \land \mathcal{A}_Q(q) \subseteq \mathcal{A}_T(t) \land \mathcal{A}_Q(q') \subseteq \mathcal{A}_T(t') \land \mathcal{B}_Q(e_Q) \subseteq \mathcal{B}_T(e_T)$}{
            $Cand(e) := Cand(e) \cup \{e_T\}$\;
        }
    }
}
\uElseIf{$f(q) \neq undefined \land f(q') \neq undefined$}{
    \ForEach{$e_T=(f(q),f(q')) \in E_T$}{
        \uIf{$\pi_T(e_T)=\pi_Q(e_Q) \land \mathcal{B}_Q(e_Q) \subseteq \mathcal{B}_T(e_T)$}{
            $condCheck :=$ CheckEdgeBreakCond($\mathcal{BC}_E,e_Q,e_T,g$)\;
            \uIf{$condCheck$ = $true$}{
                $Cand(e) := Cand(e) \cup \{e_T\}$\;
            }
        }
    }
}
\uElseIf{$f(q) \neq undefined$}{
    \ForEach{$e_T=(f(q),t') \in Dom(q,q')$}{
        \uIf{$\pi_T(e_T)=\pi_Q(e_Q) \land \mathcal{A}_Q(q') \subseteq \mathcal{A}_T(t') \land \mathcal{B}_Q(e_Q) \subseteq \mathcal{B}_T(e_T)$}{
            $condCheck :=$ CheckNodeBreakCond($\mathcal{BC}_N,q',t',f$)\;
            \uIf{$condCheck$ = $true$}{
                $Cand(e) := Cand(e) \cup \{e_T\}$\;
            }
        }
    }
}
\Else{
    \ForEach{$e_T=(t,f(q')) \in Dom(q,q')$}{
        \uIf{$\pi_T(e_T)=\pi_Q(e_Q) \land \mathcal{A}_Q(q) \subseteq \mathcal{A}_T(t) \land \mathcal{B}_Q(e_Q) \subseteq \mathcal{B}_T(e_T)$}{
            $condCheck :=$ CheckNodeBreakCond($\mathcal{BC}_N,q,t,f$)\;
            \uIf{$condCheck$ = $true$}{
                $Cand(e) := Cand(e) \cup \{e_T\}$\;
            }
        }
    }
}
\KwRet{$Cand$}
\caption{\textsc{FindCandidates}($Q,e,T,\mathcal{BC}_N,\mathcal{BC}_E,Dom,f,g$)}
\label{FindCandidates_Algo}
\end{algorithm}

The content of $Cand(e_Q)$ depends on whether $q$ and/or $q'$ have already been mapped or not. This leads to the four cases illustrated in the pseudocode of Algorithm \ref{FindCandidates_Algo}. In all these cases, before adding a target edge $e_T$ to $Cand(e_Q)$, we must ensure that $e_T$ is in the compatibility domain of the pair $(q,q')$ and matches the type and properties of $e_Q$. If one or both endpoints of $e_T$ are still unmapped (lines 1-4, lines 11-16 and lines 18-22) we also need to check that such nodes match the labels and the properties of corresponding $e_Q$'s endpoints. Moreover, except for the matching of the first target edge (lines 1-4), the algorithm also verifies if adding $e_T$ to the match violates one or more symmetry breaking conditions involving $e_T$ or one of its endpoints not yet mapped. For example, if only $q$ has been mapped (lines 11-16), then $Cand(e_Q)$ is the set of all possible target edges $e_T=(f(q),t')$ in the compatibility domain of pair $(q,q')$ such that: i) $e_T$ has the same type of $e_Q$ and contains all its properties, ii) $t'$ contains all the properties of $q'$ and iii) symmetry breaking conditions on $t'$ are satisfied.

Checking breaking conditions for a candidate target node $t$ to be mapped with a query node $q$ (lines 14 and 20) involves verifying if $id(f(q'))<id(t)$ for all query nodes $q'$ such that i) there exists a breaking condition $q' \prec q$ and ii) $q'$ has already been mapped. Likewise, checking breaking conditions for a candidate target edge $e_T$ to be mapped with a query edge $e_Q$ (line 8) implies verifying if $id(g(e'_Q))<id(e_T)$ for all query edges $e'_Q$ such that i) there exists a breaking condition $e'_Q \prec e_Q$ and ii) $e'_Q$ has already been mapped.

\subsection{Handling queries with WHERE clause}

MultiGraphMatch can be extended to handle conditions on labels, types and properties expressed by the WHERE clause in CYPHER language.

Before processing a query with a WHERE clause, the logical proposition $P$ expressed by the clause is first transformed into Disjunctive Normal Form (DNF), so that $P$ will have the following standard representation:

\begin{equation*}
P = (C_{11} \land C_{12} \cdots C_{1k_{1}}) \lor (C_{21} \land C_{22} \cdots C_{2k_{2}}) \lor \cdots \lor (C_{m1} \land C_{m2} \cdots C_{mk_{m}})
\end{equation*}

where $C_{ij}$ are boolean conditions involving node labels, edge types, node properties or edge properties, optionally preceded by the NOT operator that, if present, negates the value of the condition.

Conditions of the logical proposition $P$ in DNF form are then processed by MultiGraphMatch, depending on the structure of $P$. 

Let us first consider the case where $P$ does not contain OR operators. This means that $P$ is formed by a chain of boolean conditions linked by an AND operator, so they must all be satisfied in a matching subgraph.

A boolean condition of the form $A.x$ $op$ $c$, where $A$ is a node or an edge, $x$ is a label, a type or a property, $op$ is a relational operator and $c$ is a constant can be used as additional constraints to build compatibility domains of query nodes. The net result will be to allow only target edges that satisfy the conditions. 

By contrast, conditions of the form $A.x$ $op$ $B.y$, where $B$ is a second node or edge and $y$ is a label, a type or a property can be verified only at matching time, during the search for candidates (Algorithm \ref{FindCandidates_Algo}). 
In particular, the condition holds when either $A$ or $B$ (or both) are matched. Verification consists of checking whether the type or the properties of candidate edges (and optionally their endpoints) match the type and the properties of the matching query edge (lines 3, 7, 13, 19).

If the logical proposition $P$ contains OR operators, MultiGraphMatch first splits the query into several queries that have the same MATCH clause and different WHERE clauses. More formally, let $Q$ a query and $P = P_1 \lor P_2 \lor \cdots \lor P_k$ the logical proposition expressed by the WHERE clause of $Q$, where $P_1, P_2, ..., P_k$ are propositions containing only AND operators. MultiGraphMatch creates $k$ different queries $Q_1, Q_2, ..., Q_k$, where query $Q_i$ (with $1 \leq i \leq k$) has the same MATCH clause as $Q$ and the WHERE clause of $Q_i$ has the proposition $P_i$. Each query $Q_i$ contains only propositions linked by AND operators, so it can be solved as in the case where there are no OR operators. Finally, the union of the sets of solutions returned by each query is computed to find the set of solutions for $Q$.

\subsection{Complexity analysis}
\label{complexitySec}

In this section, we analyze the time and space complexity of MultiGraphMatch.

We start by evaluating the time complexity of each step of the algorithm. Let $Q=(V_Q,E_Q,\mathcal{L}_Q,\sigma_Q,$ $\mathcal{T}_Q,\pi_Q,\mathcal{A}_Q,\mathcal{B}_Q)$ the query and $T=(V_T,E_T,\mathcal{L}_T,\sigma_T,\mathcal{T}_T,\pi_T,$ $\mathcal{A}_T,\mathcal{B}_T)$ the target.

Indexing requires scanning all nodes and edges of the target multigraph. To build the node label graph, nodes are read one at a time and inserted into the graph, following a path from the root to one of the graph vertices.

Let $|\mathcal{L}_T|=l_T$. In the worst case, the node to insert has $l_T$ labels and the length of the traversed path is $l_T$. Therefore, the time required to build the node label graph on the target graph is $O(n_T \times l_T)$, where $n_T=|V_T|$. The edge types map and the edge properties table are built reading edges and associated type and properties one at a time. If the two data structures are implemented using hash maps and hash tables, accessing and updating them can be done in constant time, so building the edge types map and the edge properties table requires time $O(e_T)$, where $e_T=|E_T|$. To create the bit matrix we need to scan all target nodes and edges, with relative labels and types, so the time required to build the matrix is $O(n_T+e_T)$.

Computation of symmetry breaking conditions is related to the calculation of query automorphisms, which is a special case of the graph isomorphism problem, for which no polynomial algorithm is currently known (though the problem is not even known to be NP-complete). In our system, automorphisms are calculated using the NAUTY algorithm \cite{mckay2014practical}, which requires in the worst case $O(\exp(n_Q))$ \cite{miyazaki1997complexity}, where $n_Q=|V_Q|$ is the number of query nodes.

Compatibility domains are computed by matching the rows of the bit matrices associated to the target and the query and comparing the type-dependent in- and out-degrees of endpoints of the matched pairs of nodes. Building the bit matrix for the query requires $O(e_Q)$, where $e_Q=|E_Q|$.  The worst case holds when there is at most one edge between pairs of nodes in both the query and the target. In that case, the bit matrices of the query and the target have $e_Q$ and $e_T$ rows, respectively. Therefore, matching the rows of the two bit matrices costs $O(e_Q \times e_T)$. For each pair of matched rows, checking type-dependent degrees costs $O(t_T)$, where $t_T=|\mathcal{T}_T|$. So, the total time required to compute compatibility domains is $O(e_Q \times e_T \times t_T)$.

Calculating the processing order of query edges for the match requires computing at each step the $CF$ and the $Sc$ scores for every pair of connected query nodes that have not been added yet to the partial ordering. The number of connected pairs is $O(e_Q)$, so the two scores are computed $O(e_Q^2)$ times. In the worst case, calculation of the $Sc$ score for a pair $(q_i,q_j)$ also requires the Jaccard similarity between the neighborhoods of $q_i$ and $q_j$. Since a query node can have at most $n_Q-1$ neighbors, calculating $Sc$ costs $O(n_Q)$. Therefore, the total time required to compute the processing order is $O(e_Q^2 \times n_Q)$.

The computational complexity of the matching step mainly depends on the total number of candidate pairs of query and target edges that are examined during the process. In the worst case, the compatibility domain of each query edge has $e_T$ elements. Therefore, the set of candidate nodes for the initial query edge has at most $e_T$ edges. Once a new pair of edges has been matched, the set of candidate target edges for the next query edge in the ordering has at most $d-1$ edges, where $d$ is the maximum node degree in the target. In fact, except for the initial set of candidates, the next sets of candidates contain target edges having at least one node in common with already matched edges. By summing up, the total number of examined candidate pairs is at most $e_T+e_T(d-1)+e_T(d-1)^2+ ... +e_T(d-1)^{e_Q}$ which is bounded by $O(e_T \times d^{e_Q})$. For a given candidate pair of edges, breaking conditions are checked in $O(n_Q)$ time, while comparing node and edge properties of the two edges costs $O(p_{n_T}+p_{e_T})$, where $p_{n_T}=|\mathcal{A}_T|$ and $p_{e_T}=|\mathcal{B}_T|$. If the algorithm finds a match with a candidate we need to update the current partial match and the latter operation requires constant time. On the other hand, performing a step of backtracking also requires constant time operations. So, the total time needed to perform matching is $O(e_T \times d^{e_Q} \times (n_Q+{pn}_T+{pe}_T))$, which is bounded by $O(n_Q e_T d^{e_Q})$.

By summing the time complexities of each step of MultiGraphMatch, we get $O(\exp(n_Q) + e_Q e_T t_T + {e_Q}^2 n_Q + n_Q e_T d^{e_Q})$. In practice, the query is much smaller than the target, in which case the worst case time complexity of our algorithm is  $O(n_Q e_T d^{e_Q})$. While this worst case complexity is exponential in the degree, in practice, the complexity is much smaller because MultiGraphMatch processes and filters based on each node pair of the query graph, using the bit matrix.

Regarding the spatial complexity of MultiGraphMatch, most of the memory is used to store the four indexing data structures on the target. In the node label graph each node with associated properties is stored only once, so this data structure costs $O(n_T \times {pn}_T)$. In the edge type map each edge is stored in only one of the inner hash maps, so this index requires $O(e_T)$ space. The edge properties table costs $O(e_T \times {pe}_T)$ space. The bit matrix of the target contains a row for each ordered pair of connected nodes. The number of such pairs is at most $O(e_T)$, while the number of columns of the bit matrix is $2l_T+2$. So, the bit matrix costs $O(e_T \times l_T)$ space. Summing up, the total space required to store the indexing data structures for the target is $O(n_T \times pn_T+e_T \times (pe_T+l_T))$. In a large network, we can reasonably assume that the number of labels and properties is negligible compared to the number of nodes and edges of the target, so the required space is $O(e_T)$. Additional data structures used by the algorithm in the next steps include the set of symmetry breaking conditions for the query ($O(n_Q^2+e_Q^2)$ space), the bit matrix for the query ($O(e_Q l_Q)$ space, where $l_Q=|\mathcal{L}_Q|$), the set of compatibility domains for each query edge ($O(e_Q e_T)$ space in the worst case), the node and edge mappings ($O(n_Q)$ and $O(e_Q)$ space, respectively) and the set of candidate target edges for the matching for each query edge (which costs $O(e_Q e_T)$ in the worst case). Therefore, the total spatial complexity of MultiGraphMatch is $O(e_Q \times e_T)$.

\section{Experimental Results}
\label{experiments}
We performed a series of experiments on both synthetic and real networks to compare the performance of the MultiGraphMatch algorithm with state-of-the-art methods: Neo4J (version 4.2) (\url{https://neo4j.com/}), Memgraph (\url{https://memgraph.com/}) and SuMGra \cite{ingalalli2016sumgra}. 

According to recent work \cite{monteiro2023experimental}, Neo4J is the most efficient graph database system. Furthermore, Neo4J provides full support for queries in multigraphs, unlike many other systems. In our experiments, we also included Memgraph, because it compared favorably with Neo4J (version 3). Moorman's algorithm \cite{moorman2021subgraph} was excluded from the comparison due to its high memory requirements even for networks of medium size in our tests. We also excluded \cite{anwar2022subgraph} and FG$q_{t}$-Match \cite{sun2021subgraph} from our analysis, because we could not obtain the software from the authors. 
 
We performed additional experiments to evaluate i) the impact of newly introduced features, such as bit matrix and processing ordering of query edges, on the performance of MultiGraphMatch and ii) the scalability of our algorithm.

All experiments entail counting all the occurrences of a given query in a target and were performed on an Intel(R) Xeon(R) Gold 6132 CPU 2.60GHz with 400 GB of RAM. MultiGraphMatch has been implemented in Java and is available at\\ \href{https://github.com/Anto188bas/MultiGraphMatch.git}{https://github.com/Anto188bas/MultiGraphMatch.git}, together with the datasets used for the experiments.

\subsection{Experiments on synthetic networks}

\subsubsection{Data description}
\label{syntheticdatasets}

We built a dataset of labeled synthetic networks, generated according to the Barabasi-Albert model \cite{barabasi1999emergence}, implemented within the snap tool \cite{leskovec2016snap}. All networks contain 10,000 nodes and 1 million edges. Each node in the networks is associated with a single label, randomly sampled from a set of possible labels according either to a power law with exponent $-1.2$ or to a uniform distribution. An analogous algorithm is used to label each edge of the networks. The number of distinct node labels and the number of distinct edge labels in the synthetic networks are either 2 or 10.

This results in a final set of eight different synthetic networks:

\begin{itemize}
\item \textsc{Uniform(2,2)}: a network with 2 distinct node labels and 2 distinct edge labels sampled from a uniform distribution;
\item \textsc{Uniform(2,10)}: a network with 2 distinct node labels and 10 distinct edge labels sampled from a uniform distribution;
\item \textsc{Uniform(10,2)}: a network with 10 distinct node labels and 2 distinct edge labels sampled from a uniform distribution;
\item \textsc{Uniform(10,10)}: a network with 10 distinct node labels and 10 distinct edge labels sampled from a uniform distribution;
\item \textsc{PowerLaw(2,2)}: a network with 2 distinct node labels and 2 distinct edge labels sampled from a power-law distribution;
\item \textsc{PowerLaw(2,10)}: a network with 2 distinct node labels and 10 distinct edge labels sampled from a power-law distribution;
\item \textsc{PowerLaw(10,2)}: a network with 10 distinct node labels and 2 distinct edge labels sampled from a power-law distribution;
\item \textsc{PowerLaw(10,10)}: a network with 10 distinct node labels and 10 distinct edge labels sampled from a power-law distribution;
\end{itemize}

\subsubsection{Performance comparison}
\label{performanceSynthetic}

We first compared MultiGraphMatch, SuMGra, Memgraph and Neo4j on the dataset of synthetic networks.

To this end, from each network, we randomly extracted 600 query graphs with different number of nodes (from 3 to 8) and densities (from 0.25 to 1), where density is defined as the fraction of query node pairs connected by at least one edge. Queries were extracted as follows. Let $k$ and $d$ be the desired number of nodes and density of the extracted subgraph, respectively. First, we perform a random walk of length $k$ and select all the nodes and edges visited during the walk. Next, we randomly add edges that: i) connect nodes belonging to the random walk and ii) have not been visited before. This is done until the extracted subgraph has the desired density. If there are not enough edges to add, the extraction restarts from a new node. 

Experimental results are summarized in Table \ref{resSynthetic}. For each synthetic network and for each algorithm, the table reports the number of queries completed before the timeout of 30 minutes (1,800 seconds), the mean running time and the 90\% confidence interval of the mean time. 

Reported results refer only to queries completed by at least one algorithm before the timeout, with a running time of 1,800 seconds assigned to each uncompleted query.

\begin{table}
\footnotesize
    \centering
    \begin{tabular}{ccccc}
        \toprule
        \multirow{2}{*}{\bf Network} & \multirow{2}{*}{\bf Algorithm} & {\bf Completed} & \multirow{2}{*}{\bf Mean time (secs)} & {\bf Confidence} \\ 
        & & {\bf queries} & & {\bf interval} \\
        \midrule       
        \multirow{4}{*}{\textsc{Uniform(2,2)}} & MGM & 340 & {\bf 223.83} & $[192.07, 259.50]$ \\
        & Neo4J & 267 & 640.71 & $[565.50, 721.99]$ \\
        & SuMGra & 338 & 292.94 & $[252.03, 337.70]$ \\
        & Mem & 145 & 1188.98 & $[1104.87, 1271.69]$\\ \midrule
        \multirow{4}{*}{\textsc{Uniform(2,10)}} & MGM & 556 & 17.62 & $[7.74, 32.55]$ \\
        & Neo4J & 555 & 90.86 & $[69.96, 115.34]$ \\
        & SuMGra & 559 & {\bf 16.06} & $[8.42, 27.05]$ \\
        & Mem & 316 & 912.05 & $[843.93, 981.78]$\\ \midrule
        \multirow{4}{*}{\textsc{Uniform(10,2)}} & MGM & 583 & {\bf 21.04} & $[11.85, 33.63]$ \\
        & Neo4J & 571 & 74.79 & $[52.14, 101.08]$ \\
        & SuMGra & 575 & 66.85 & $[46.52, 91.03]$ \\
        & Mem & 313 & 909.59 & $[841.84, 978.93]$\\ \midrule
        \multirow{4}{*}{\textsc{Uniform(10,10)}} & MGM & 600 & {\bf 0.03} & $[0.02, 0.04]$ \\
        & Neo4J & 600 & 0.73 & $[0.64, 0.85]$ \\
        & SuMGra & 600 & 0.15 & $[0.09, 0.23]$ \\
        & Mem & 600 & 68.95 & $[49.79, 90.97]$\\ \midrule
        \multirow{4}{*}{\textsc{PowerLaw(2,2)}} & MGM & 308 & {\bf 287.76} & $[240.48, 336.69]$ \\
        & Neo4J & 206 & 759.19 & $[671.98, 851.37]$ \\
        & SuMGra & 293 & 406.69 & $[344.86, 468.07]$ \\
        & Mem & 151 & 1075.21 & $[984.56, 1166.70]$\\ \midrule
        \multirow{4}{*}{\textsc{PowerLaw(2,10)}} & MGM & 564 & {\bf 38.15} & $[23.84, 56.45]$ \\
        & Neo4J & 525 & 199.42 & $[160.47, 240.92]$ \\
        & SuMGra & 557 & 53.14 & $[35.05, 75.89]$ \\
        & Mem & 280 & 1001.44 & $[930.62, 1070.48]$\\ \midrule
        \multirow{4}{*}{\textsc{PowerLaw(10,2)}} & MGM & 568 & {\bf 46.40} & $[32.51, 63.98]$ \\
        & Neo4J & 537 & 172.15 & $[136.75, 212.65]$ \\
        & SuMGra & 549 & 124.13 & $[94.91, 156.71]$ \\
        & Mem & 299 & 966.42 & $[896.23, 1035.33]$\\ \midrule
        \multirow{4}{*}{\textsc{PowerLaw(10,10)}} & MGM & 600 & {\bf 1.92} & $[0.41, 5.22]$ \\
        & Neo4J & 600 & 11.17 & $[4.20, 22.05]$ \\
        & SuMGra & 600 & 9.63 & $[2.67, 20.81]$ \\
        & Mem & 600 & 219.12 & $[178.57, 261.53]$\\       
        \bottomrule
    \end{tabular}
    \caption{Performance of MultiGraphMatch (MGM), SuMGra, Neo4J and Memgraph (Mem) on a dataset of queries run on synthetic networks. The table reports, for each synthetic network and for each algorithm, the number of queries completed before the timeout (30 minutes), the mean running time (in seconds) and the  90\% confidence interval. Uncompleted queries are considered to take 30 minutes. MultiGraphMatch is better for all synthetic networks except \textsc{Uniform(2,10)} where SuMGra is slightly faster.}
    \label{resSynthetic}
\end{table}

\begin{figure}
    \centering
    \includegraphics[width=11cm]{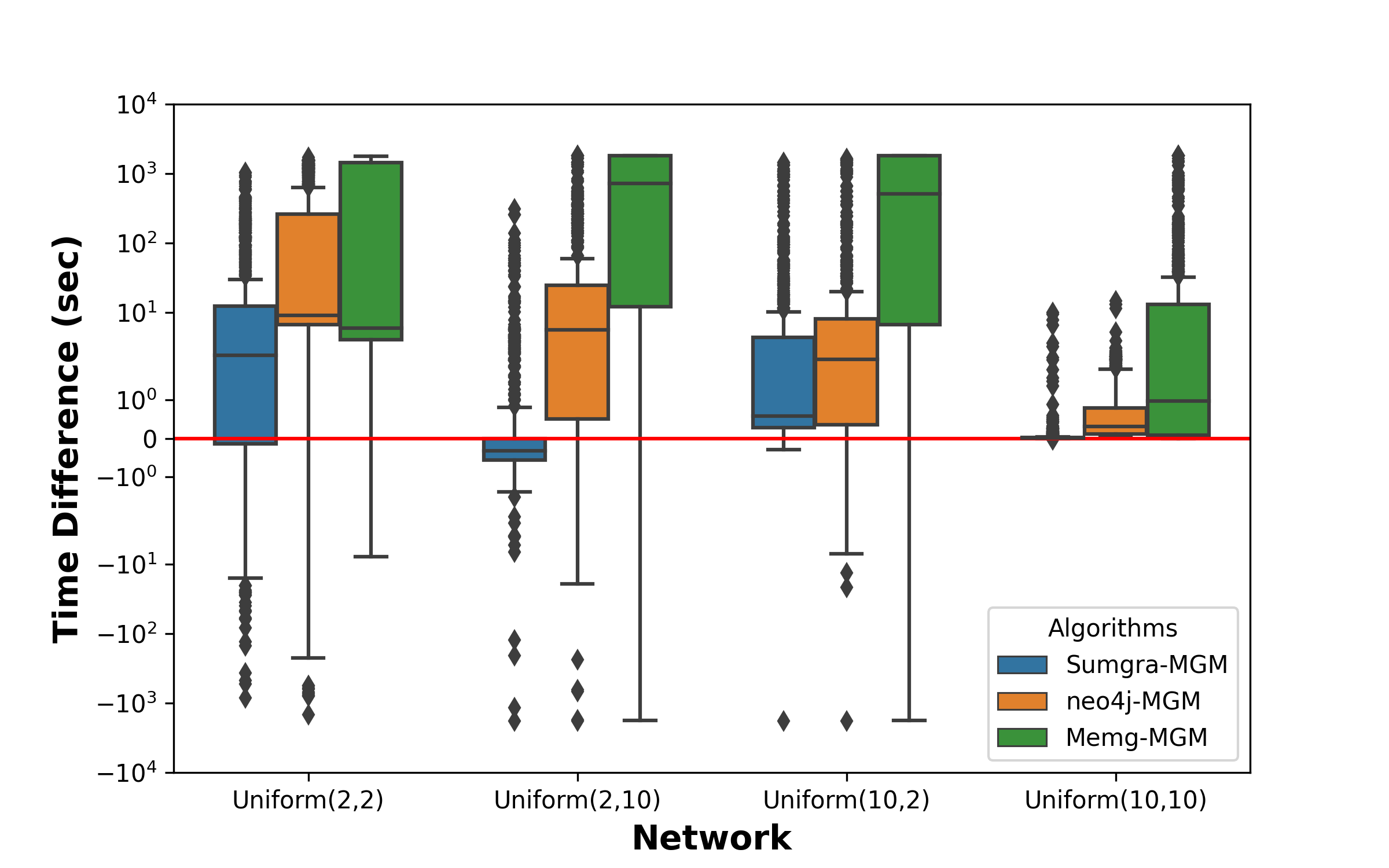}
    \caption{Boxplots of the query-by-query running time differences (in seconds) between MultiGraphMatch (MGM) and the other compared tools SuMGra, Neo4J and Memgraph (Mem) for the dataset of queries run on the uniform synthetic networks. If the mean value of another system $S$ lies on the zero line, then its mean is the same as MGM. If above, then $S$ is slower (in the mean) than MGM. The boxplots show that  the mean value of MultiGraphMatch is the lowest in every case except for \textsc{Uniform(2,10)} where SuMGra enjoys a small but not significant advantage.}
    \label{boxplotSyntheticUniform}
    \Description[Running time differences in random networks]{Boxplots of the query-by-query running time differences (in seconds) between MultiGraphMatch (MGM) and the other compared tools SuMGra, Neo4J and Memgraph (Mem) for the dataset of queries run on the uniform synthetic networks}
\end{figure}

\begin{figure}
    \centering
        \includegraphics[width=11cm]{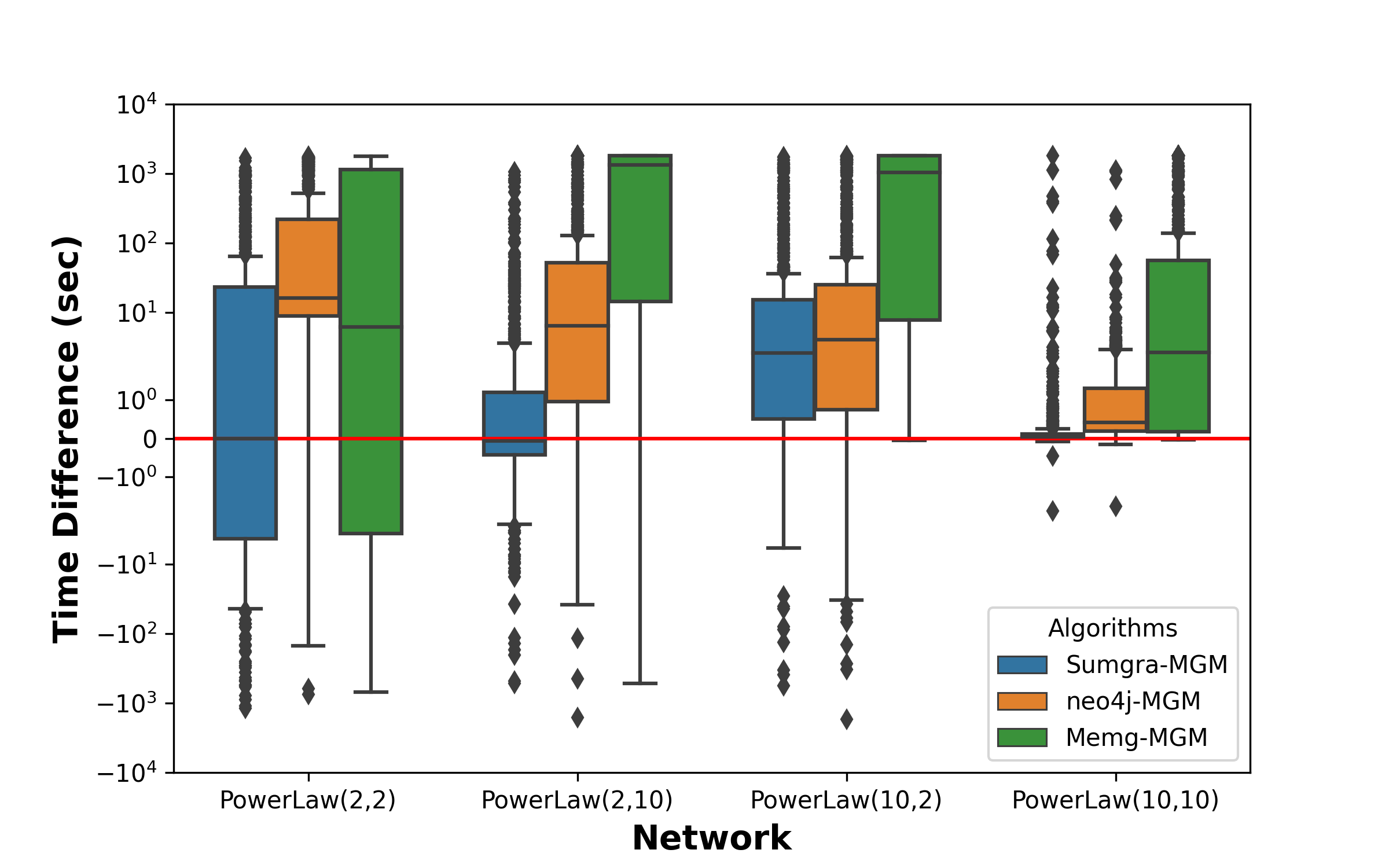}
    \caption{Boxplots of the query-by-query running time differences (in seconds) between MultiGraphMatch (MGM) and the other compared tools SuMGra, Neo4J and Memgraph (Mem) for the dataset of queries run on power-law synthetic networks. MultiGraphMatch is as fast as or significantly faster than the other methods.}
    \label{boxplotSyntheticFractal}
    \Description[Running time differences in power-law networks]{Boxplots of the query-by-query running time differences (in seconds) between MultiGraphMatch (MGM) and the other compared tools SuMGra, Neo4J and Memgraph (Mem) for the dataset of queries run on power-law synthetic networks}
\end{figure}

Figures \ref{boxplotSyntheticUniform} and \ref{boxplotSyntheticFractal} illustrate boxplots of the distribution of the query-by-query running time differences between MultiGraphMatch and the other compared algorithms in each synthetic network. To build these plots, we considered only queries completed by at least one tool before the timeout and we assigned a running time of 1,800 seconds for each uncompleted query. In Supplementary Table \ref{resSuppSynthetic} we also list the median, the 90\% confidence interval and the p-value of significance of the running time differences between MultiGraphMatch and the other algorithms. The p-values and confidence intervals were computed non-parametrically \cite{katari2021statistics}.

Results clearly show that MultiGraphMatch is the fastest algorithm (with p-value < 0.05) except for the network \textsc{Uniform(2,10)}, where SuMGra is slightly faster but with a non-significant p-value ($0.35$). The advantage of MultiGraphMatch is overall higher in power-law networks. Moreover, except for the network \textsc{Uniform(2,10)}, MultiGraphMatch completed the most queries before the timeout.

\subsubsection{Ablation Tests}
\label{ablationExp}

MultiGraphMatch introduces (i) a bit matrix data structure (see Section \ref{IndexingSec}) for submultigraph matching and (ii) a novel ordering scheme for processing query edges (see Section \ref{ordering}). We ran a series of ablation tests on a subset of synthetic networks to evaluate the impact of each of these features on the performance of our algorithm.

%\color{red}

We compared the full MultiGraphMatch (abbreviated as MGM) with six simplified versions of the algorithm, each of them missing one or more of our heuristics. To test the impact of the bit matrix on performance, we considered MultiGraphMatch with no Bit Matrix (MGM-NoBM). Concerning the processing order, we relied on the most common criteria used to order query nodes or edges in subgraph matching, i.e. node degree, label frequency and the cardinality of candidate sets \cite{bonnici2017variable}. Therefore, we considered MultiGraphMatch with Random Ordering (MGM-RO), MultiGraphMatch with Ordering based on minimization of Domain Cardinalities (MGM-DC), MultiGraphMatch with ordering based on minimization of the frequency of Edge Labels (MGM-EL) and MultiGraphMatch with Ordering based solely on the numerator of the Sc score defined in Section \ref{ordering} (MGM-DE), which is related to the degrees of edge's endpoints. As a baseline for the comparison, we considered MultiGraphMatch with no BitMatrix and Random Ordering (MGM-NoBM-RO).

%\color{black}

The experiments were performed on the complete set of queries extracted from the following synthetic networks: \textsc{Uniform(2,10)}, \textsc{Uniform(10,2)}, \textsc{PowerLaw(2,10)} and \textsc{PowerLaw(10,2)}. Also in this case, a timeout of 30 minutes was set for the execution of each query.

\begin{figure}
    \centering
    \includegraphics[width=12cm]{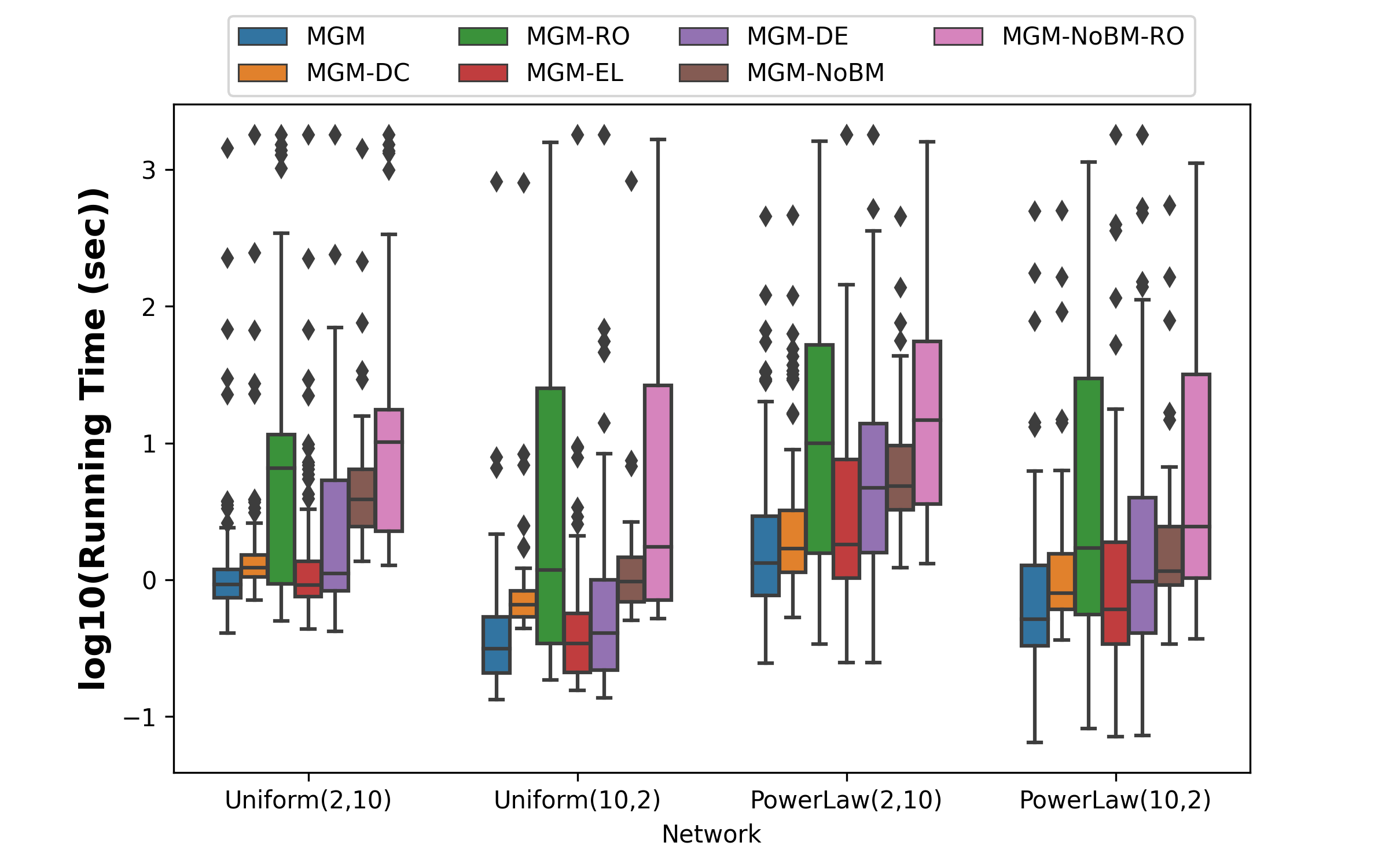}
    \caption{\textcolor{black}{Boxplots of the log of the running times (in seconds) of MultiGraphMatch (MGM), MultiGraphMatch with no Bit Matrix (MGM-NoBM), MultiGraphMatch with Random Ordering (MGM-RO), MultiGraphMatch with Ordering based on Minimization of Domain Cardinalities (MGM-DC), MultiGraphMatch with ordering based on minimization of the frequency of Edge Labels (MGM-EL), MultiGraphMatch with ordering based solely on the numerator of the Sc score defined in Section \ref{ordering} (MGM-DE) and MultiGraphMatch with no Bit Matrix and Random Ordering (MGM-NoBM-RO) on the following synthetic networks: \textsc{Uniform(2,10)}, \textsc{Uniform(10,2)}, \textsc{PowerLaw(2,10)} and \textsc{PowerLaw(10,2)}. Ordering and the bit matrix significantly improve performance, both individually and collectively. All values are relative to a 0 point which is the mean of the matching time of MGM on \textsc{Uniform(2,10)}. Lower is better. Thus, all these heuristics are helpful.}}
    \label{boxplotAblationTests}
    \Description[Ablation tests on synthetic networks]{Boxplots of the log of the running times (in seconds) of MultiGraphMatch (MGM), MultiGraphMatch with no Bit Matrix (MGM-NoBM), MultiGraphMatch with Random Ordering (MGM-RO), MultiGraphMatch with Ordering based on Minimization of Domain Cardinalities (MGM-DC), MultiGraphMatch with ordering based on minimization of the frequency of Edge Labels (MGM-EL), MultiGraphMatch with ordering based solely on the numerator of the Sc score defined in Section \ref{ordering} (MGM-DE) and MultiGraphMatch with no Bit Matrix and Random Ordering (MGM-NoBM-RO) on the following synthetic networks: \textsc{Uniform(2,10)}, \textsc{Uniform(10,2)}, \textsc{PowerLaw(2,10)} and \textsc{PowerLaw(10,2)}}
\end{figure}

Fig. \ref{boxplotAblationTests} shows box plots of the running times of the compared versions of MultiGraphMatch on each tested synthetic network. 

%\color{red}

In Supplementary Table \ref{resSuppAblation} we also report the median, the 90\% confidence interval and the p-value of significance of the query-by-query running time differences between the full MultiGraphMatch algorithm and its simplified versions. The p-values were computed non-parametrically \cite{katari2021statistics}.

Results show that the full algorithm is faster than its simplified versions, implying that both the bit matrix and the ordering are important for the performance of MultiGraphMatch. 

The bit matrix improves on candidate detection and computation of compatibility domains thanks to the bit signature representation of the types of all edges connecting two nodes and the labels of the source and destination nodes of such edges. This representation is also more memory-efficient and less time-consuming (high performance of bitwise operations) than a solution that uses a vector of vectors or a map of maps to identify compatible edges. Ordering query edges greatly speeds up matching. An incorrect ordering of query edges may increase the number of backtracking operations and, consequently, the matching time. 

Regarding ordering, we found that the complete scoring function worked much better than the scoring functions based only on either minimization of domain cardinalities (MGM-DC) or frequency of edge labels (MGM-EL) for tree queries. For such queries, both MGM-DC and MGM-EL often start matching from peripheral edges. Compared to the scoring scheme based only on edge’s endpoint degrees (MGM-DE), the full scoring function worked generally better, especially in the two power-law networks.

%\color{black}

\subsection{Experiments on real networks}

\subsubsection{Data description}\label{realdatasets}

The performance of the various matching algorithms was also evaluated on three real networks of different sizes: \textsc{imdb}, \textsc{panama} and \textsc{vector}. Tab. \ref{tab:real} summarizes their main features.

\begin{table}
    \footnotesize
    \centering
    \begin{tabular}{lcccc}
        \toprule
        \textbf{Network} & $|V|$ & $|E|$ & $|\mathcal{L}|$ & $|\mathcal{T}|$  \\ \midrule
        IMDB    & 9,930,907 & 37,874,660 & 67 & 10 \\
        PANAMA  & 1,908,466 & 3,142,523  & 5  & 13 \\
        VECTOR  & 19,568    & 16,027,761 & 6  &  2 \\
        \bottomrule
    \end{tabular}
    \caption{Main features of the three real networks used in the experiments. For each network we report the number of nodes $|V|$, the number of edges $|E|$, the number of distinct node labels $|\mathcal{L}|$ and the number of distinct edge types $|\mathcal{T}|$.}
    \label{tab:real}
\end{table}

\textsc{imdb} is a network extracted from the Internet Movie Database (IMDB)\footnote{https://www.imdb.com/} that provides information about millions of movies, television, home videos, video games, and online streaming content programs, including cast and crew. The nodes of \textsc{IMDB} are people working in the show business (actors, producers, composers, writers, etc.) and movies, while edges denote different types of relationship between actors and movies (e.g. an actor who played in a movie or a director who directed a movie). Person nodes are annotated with labels denoting the main professions of a person. Movie nodes have labels denoting the genre of a movie. Properties associated to people are the name of a person, the year of birth and the year of death (if not alive), while properties linked to a movie are its original title and the year of production. The \textsc{imdb} network was built starting from information stored in flat text files publicly available at \url{https://imdb-api.com/}.

\textsc{panama} is a multigraph extracted from the Panama Papers, a set of 11.5 million leaked documents containing detailed financial information of more than 200,000 offshore entities and their clients from its foundation in 1977 to April 2015. The documents originally belonged to the Panamanian law firm and corporate service provider Mossack Fonseca. They were leaked in 2015 by an anonymous source. The documents reveal that some of the Mossack Fonseca shell corporations were used for illegal purposes, including fraud, tax evasion, and evading international sanctions. Intermediaries (e.g. law firms), banks or trust companies with external advisors who connect the client to offshore service providers (e.g. Mossack Fonseca) support these activities. Recently, the International Consortium of Investigative Journalists (ICIJ)\footnote{https://www.icij.org} collected all this information in a freely accessible database called Panama\footnote{https://offshoreleaks.icij.org}. From this database, we extracted the directed relationship network between offshore entities, intermediaries and officers (i.e. people or companies playing a role in an offshore entity). Nodes are labeled according to their type ('intermediary', 'entity', 'officer' or 'address' of offshore companies). Edges denote the kind of relationship (e.g. 'intermediary of' links an intermediary to an entity, while 'officer of' links an officer to an entity).

\textsc{vector} \cite{barbagallo2021vector} is a Uveal Melanoma (a type of eye cancer) knowledge network containing: i) the expression of mRNA, transcription factors (TFs) and non-coding RNAs (e.g. miRNAs, lncRNAs) in several human normal tissues and cancer samples, ii) the correlation between the expression profiles of mRNAs, TFs and non-coding RNAs in normal tissues and cancer samples. Expression data are taken from The Cancer Genome Atlas (TCGA)\footnote{https://www.cancer.gov/about-nci/organization/ccg/research/structural-genomics/tcga}, a cancer genomics program that collects molecular data concerning about 20,000 primary cancer and matched normal samples spanning 33 different cancer types. Starting from the expression data, Pearson correlations between (a) miRNAs and mRNAs, (b) miRNAs and lncRNAs, and (c) lncRNAs and mRNAs were computed.

\subsubsection{Performance comparison}
\label{performanceReal}

Performance on real networks were evaluated considering two sets of queries: queries having only specified values on node labels and edge types (i.e. without WHERE clauses) and queries also having conditions on node and edge properties (i.e. with WHERE clauses). Queries with WHERE clauses were performed on all real networks, while queries without WHERE clause were tested only on \textsc{imdb} and \textsc{panama}, because \textsc{vector} is quite dense, therefore queries were not very selective, resulting in many timeouts. In addition, SuMGra cannot handle multigraphs with properties, for this reason SuMGra was run only on queries without WHERE clause. 

We randomly extracted from each real network 300 queries with WHERE clause and 400 queries without WHERE clause having different number of nodes (from 3 to 5) and density values (from 0.25 to 1). The extraction of queries was performed using the same random walk algorithm described in Section \ref{performanceSynthetic}. WHERE clauses were built by randomly imposing equality constraints (from 1 to 3) on node and edge properties.

Experimental results obtained on queries without and with WHERE clauses are summarized in Tables \ref{resRealNoWhere} and \ref{resRealWhere}, respectively. The two tables report, for each network and for each algorithm, the number of queries completed before the timeout, the mean running time and the relative 90\% confidence interval of the mean time. Reported results refer only to queries completed by at least one algorithm before the timeout, with a running time of 1,800 seconds assigned for each uncompleted query.

\begin{table}
\footnotesize
    \centering
    \begin{tabular}{ccccc}
        \toprule
        \multirow{2}{*}{\bf Network} & \multirow{2}{*}{\bf Algorithm} & {\bf Completed} & \multirow{2}{*}{\bf Mean time (secs)} & {\bf Confidence} \\ 
        & & {\bf queries} & & {\bf interval} \\
        \midrule       
        \multirow{4}{*}{\textsc{IMDB}} & MGM & 358 & {\bf 256.50} & $[163.09, 391.47]$ \\
        & Neo4J & 304 & 386.12 & $[317.88, 454.47]$ \\
        & SuMGra & 315 & 344.86 & $[283.78, 412.53]$ \\
        & Mem & 315 & 286.21 & $[225.45, 349.13]$\\ \midrule
        \multirow{4}{*}{\textsc{PANAMA}} & MGM & 162 & {\bf 31.01} & $[19.97, 46.76]$ \\
        & Neo4J & 149 & 231.06 & $[157.71, 320.34]$ \\
        & SuMGra & 161 & 115.75 & $[69.13, 174.04]$ \\
        & Mem & 128 & 481.38 & $[374.82, 596.47]$\\    
        \bottomrule
    \end{tabular}
    \caption{Performance of MultiGraphMatch (MGM), SuMGra, Neo4J and Memgraph (Mem) on a dataset of queries without WHERE clause extracted from the \textsc{imdb} and \textsc{panama} networks. The table reports, for each synthetic network and for each algorithm, the number of queries completed before the timeout (30 minutes), the mean running time (in seconds) and the relative 90\% confidence interval. Uncompleted queries are considered to take 30 minutes. MultiGraphMatch is significantly better on the two networks compared to Neo4j, SuMGra, and Memgraph.}
    \label{resRealNoWhere}
\end{table}

\begin{table}
\footnotesize
    \centering
    \begin{tabular}{ccccc}
        \toprule
        \multirow{2}{*}{\bf Network} & \multirow{2}{*}{\bf Algorithm} & {\bf Completed} & \multirow{2}{*}{\bf Mean time (secs)} & {\bf Confidence} \\ 
        & & {\bf queries} & & {\bf interval} \\
        \midrule       
        \multirow{3}{*}{\textsc{IMDB}} & MGM & 158 & {\bf 3.90} & $[1.52, 8.23]$ \\
        & Neo4J & 156 & 32.87 & $[4.29, 100.50]$ \\
        & Mem & 158 & 25.95 & $[2.36, 91.91]$\\ \midrule
        \multirow{3}{*}{\textsc{VECTOR}} & MGM & 300 & {\bf 0.05} & $[0.03, 0.07]$ \\
        & Neo4J & 300 & 0.25 & $[0.18, 0.42]$ \\
        & Mem & 300 & 83.52 & $[72.65, 94.87]$\\ \midrule
        \multirow{3}{*}{\textsc{PANAMA}} & MGM & 300 & {\bf 0.67} & $[0.46, 1.04]$ \\
        & Neo4J & 300 & 16.52 & $[5.61, 31.42]$ \\
        & Mem & 250 & 384.03 & $[310.22, 463.60]$\\    
        \bottomrule
    \end{tabular}
    \caption{Performance of MultiGraphMatch (MGM), Neo4J and Memgraph (Mem) on a dataset of queries extracted from the \textsc{imdb}, \textsc{vector} and \textsc{panama} networks having WHERE clauses. The table reports, for each synthetic network and for each algorithm, the number of queries completed before the timeout (30 minutes), the mean running time (in seconds) and the relative 90\% confidence interval. Uncompleted queries are considered to take 30 minutes. MultiGraphMatch is by far the best method.}
    \label{resRealWhere}
\end{table}

\begin{figure}
    \centering
    \includegraphics[width=11cm]{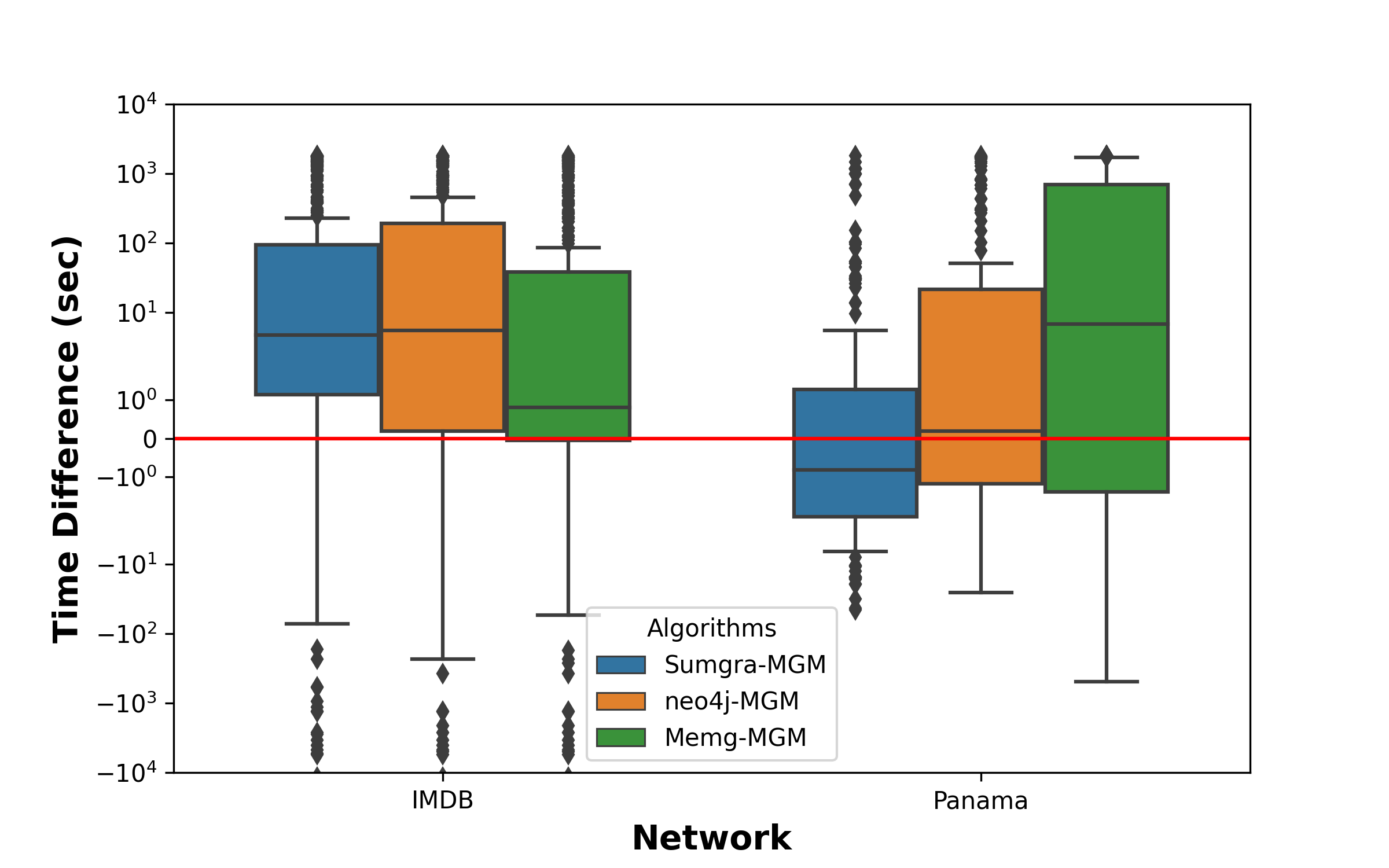}
    \caption{Boxplots of the query-by-query running time differences (in seconds) between MultiGraphMatch (MGM) and the state-of-the-art  algorithms SuMGra, Neo4J and Memgraph (Mem) for a dataset of queries without WHERE clause extracted from the \textsc{IMDB} and \textsc{PANAMA} networks.}
    \label{boxplotRealNoWhere}
    \Description[Running time differences without WHERE clause]{Boxplots of the query-by-query running time differences (in seconds) between MultiGraphMatch (MGM) and the state-of-the-art  algorithms SuMGra, Neo4J and Memgraph (Mem) for a dataset of queries without WHERE clause extracted from the \textsc{IMDB} and \textsc{PANAMA} networks}
\end{figure}

\begin{figure}
    \centering
    \includegraphics[width=11cm]{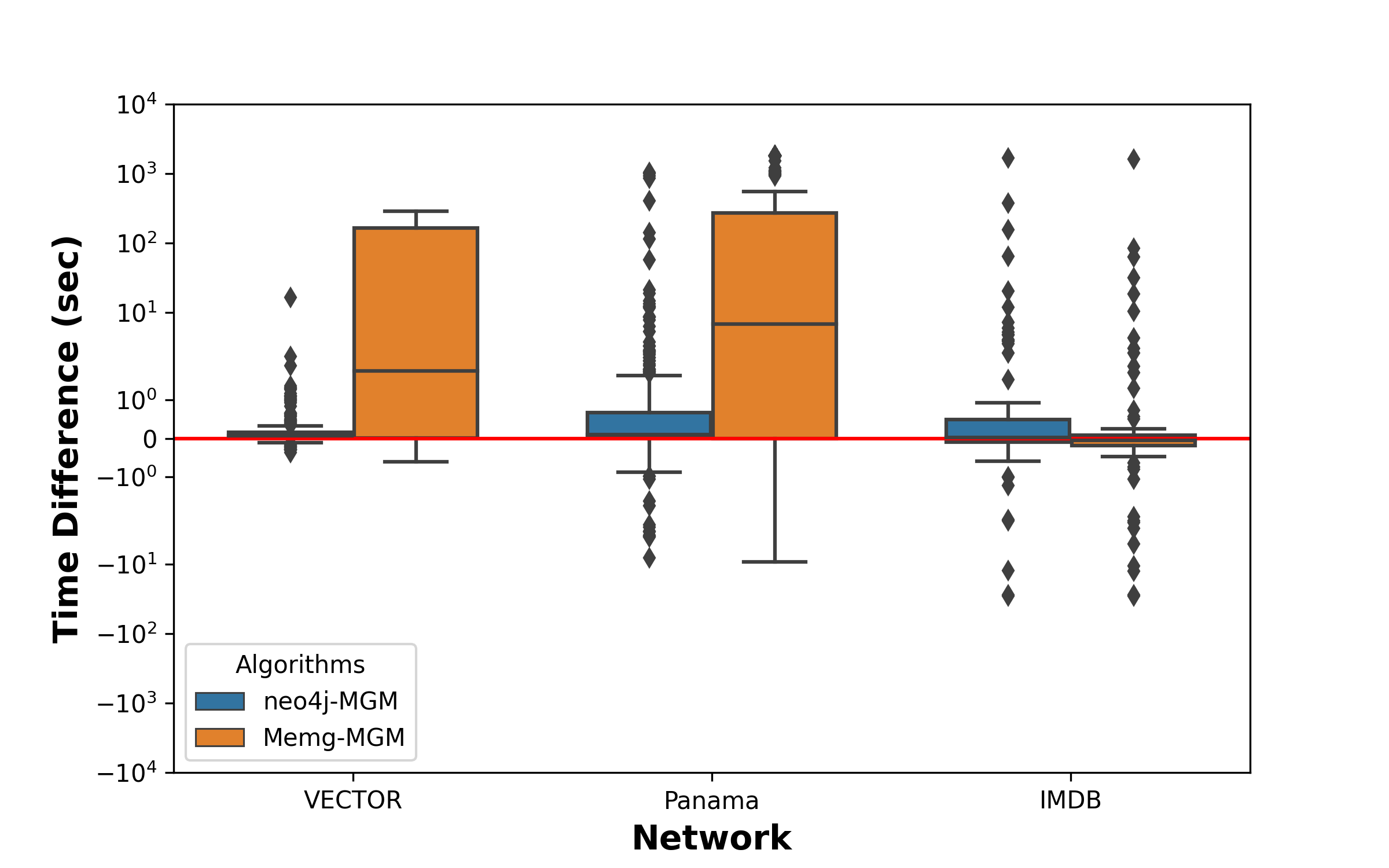}
    \caption{Boxplots of the query-by-query running time differences (in seconds) between MultiGraphMatch (MGM) and the state-of-the-art tools Neo4J and Memgraph (Mem) for the dataset of queries having WHERE clauses extracted from the \textsc{IMDB}, \textsc{PANAMA} and \textsc{VECTOR} networks.}
    \label{boxplotRealWhere}
    \Description[Running time differences with WHERE clause]{Boxplots of the query-by-query running time differences (in seconds) between MultiGraphMatch (MGM) and the state-of-the-art tools Neo4J and Memgraph (Mem) for the dataset of queries having WHERE clauses extracted from the \textsc{IMDB}, \textsc{PANAMA} and \textsc{VECTOR} networks}
\end{figure}

In Figures \ref{boxplotRealNoWhere} and \ref{boxplotRealWhere} we depict boxplots of the query-by-query running time differences between MultiGraphMatch and the other compared tools for the queries without and with WHERE clause, respectively. Boxplots were built by considering only queries completed by at least one algorithm before the timeout and assigning a running time of 1,800 seconds for each uncompleted query. In Supplementary Tables \ref{resSuppRealNoWhere} and \ref{resSuppRealWhere} we list the median, the 90\% confidence interval and the p-value of significance of the running time differences between MultiGraphMatch and the other algorithms for queries without and with WHERE clause, respectively. We computed the p-values and confidence intervals using non-parametric methods \cite{katari2021statistics}.

Results show that MultiGraphMatch significantly outperforms the other tools for queries in the absence of WHERE clauses (p-value < 0.05). For queries with WHERE clauses, MultiGraphMatch has a smaller advantage but is still the fastest tool by a significant margin. Moreover, MultiGraphMatch completed the most queries before the timeout in all scenarios.

\subsection{Scalability tests}
\label{scalabilitySec}

\subsubsection{Data description}

To evaluate the scalability of MultiGraphMatch, we used four large networks of different sizes built using the LDBC social network data generator \cite{angles2020ldbc} and six queries defined for these networks and extracted from the Labelled Subgraph Query Benchmark (LSQB)\footnote{https://github.com/ldbc/lsqb} \cite{mhedhbi2021lsqb}.

The LDBC data generator produces artificial social networks modeling the activity of a social network at different periods of time. Networks are formed by people who live in cities and countries and interact each other by establishing friendship relations and posting or replying to messages and comments in forums. People can also form groups to talk about specific topics, represented as tags. The LDBC generator creates networks that mimic the node degree distributions and attribute correlations observed in real social networks. For instance, people are more likely to travel to neighbouring countries and post messages there. The size of generated networks can be controlled through a parameter called scale factor, related to the size of produced output files. For our experiments, we used four networks with scale factors 0.1, 0.3, 1 and 3, respectively, downloaded from \url{https://github.com/ldbc/data-sets-surf-repository}. The main features of these networks are summarized in Table \ref{ldbcNetworks}.

\begin{table}
    \footnotesize
    \centering
    \begin{tabular}{lcccc}
        \toprule
        \textbf{Network} & $|V|$ & $|E|$ & $|\mathcal{L}|$ & $|\mathcal{T}|$  \\ \midrule
        LDBC\_0.1  & 432,235    & 1,806,538  & 12  & 15 \\
        LDBC\_0.3  & 1,179,535  & 5,363,741  & 12  & 15 \\
        LDBC\_1    & 3,955,790  & 18,928,261 & 12  & 15 \\
        LDBC\_3    & 11,257,915 & 55,790,601 & 12  & 15 \\
        \bottomrule
    \end{tabular}
    \caption{Main features of the 4 LDBC networks used in scalability tests. For each network we report the number of nodes, the number of edges, the number of distinct node labels and the number of distinct edge types.}
    \label{ldbcNetworks}
\end{table}

\begin{figure}
    \centering
    \includegraphics[width=11cm]{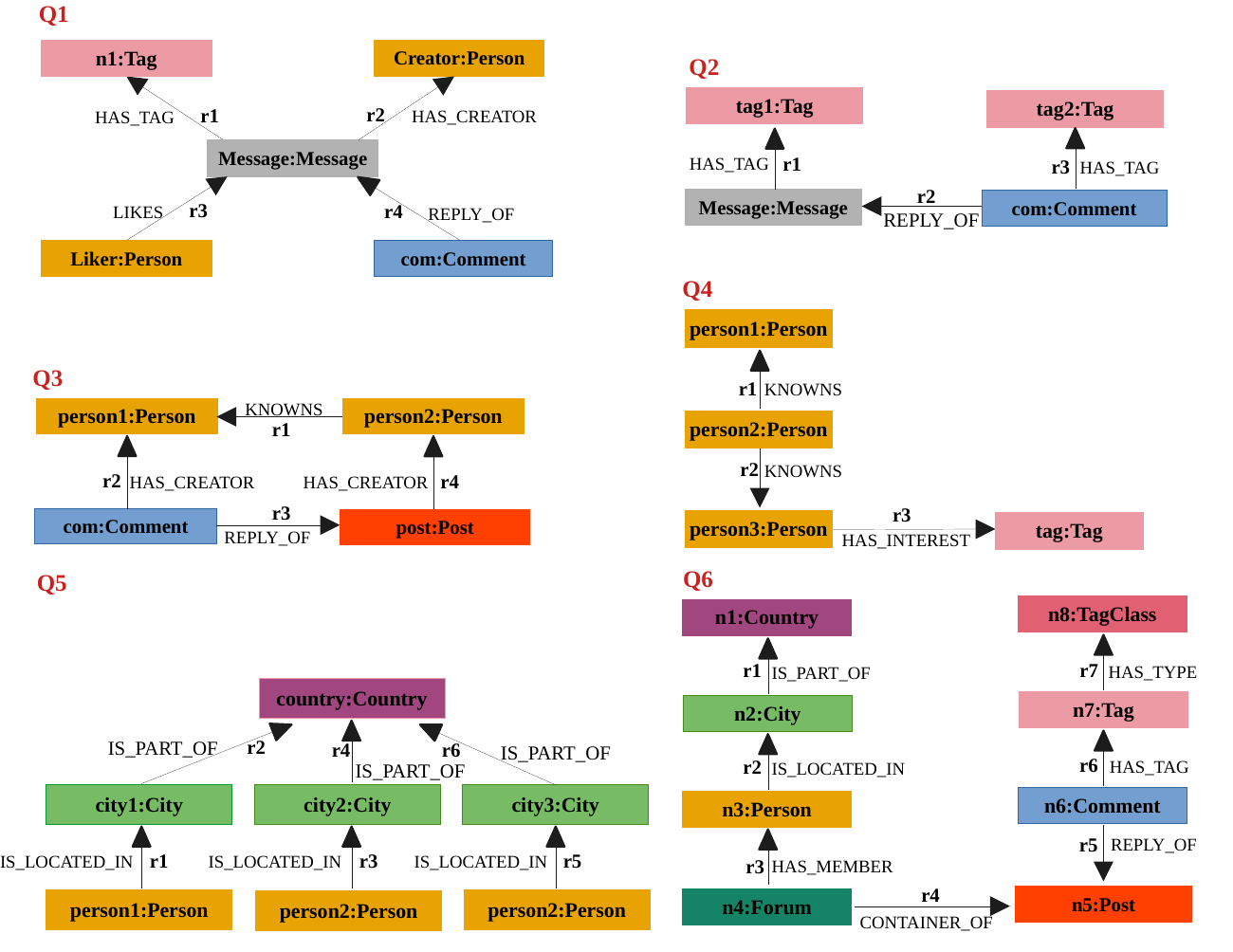}
    \caption{Queries of the LSQB benchmark used for scalability experiments.}
    \label{scalabilityQuery}
    \Description[LSQB queries]{Queries of the LSQB benchmark used for scalability experiments}
\end{figure}

Queries taken from LSQB represent patterns of different sizes and complexity typically observed in the LDBC generated networks. Queries are specifically designed to evaluate the performance of the join operation in database management systems. Figure \ref{scalabilityQuery} illustrates the six LSQB queries used for our scalability tests, named Q1, Q2, Q3, Q4, Q5 and Q6.

\subsubsection{Experimental results}

\begin{figure}
    \centering
    \includegraphics[width=9cm]{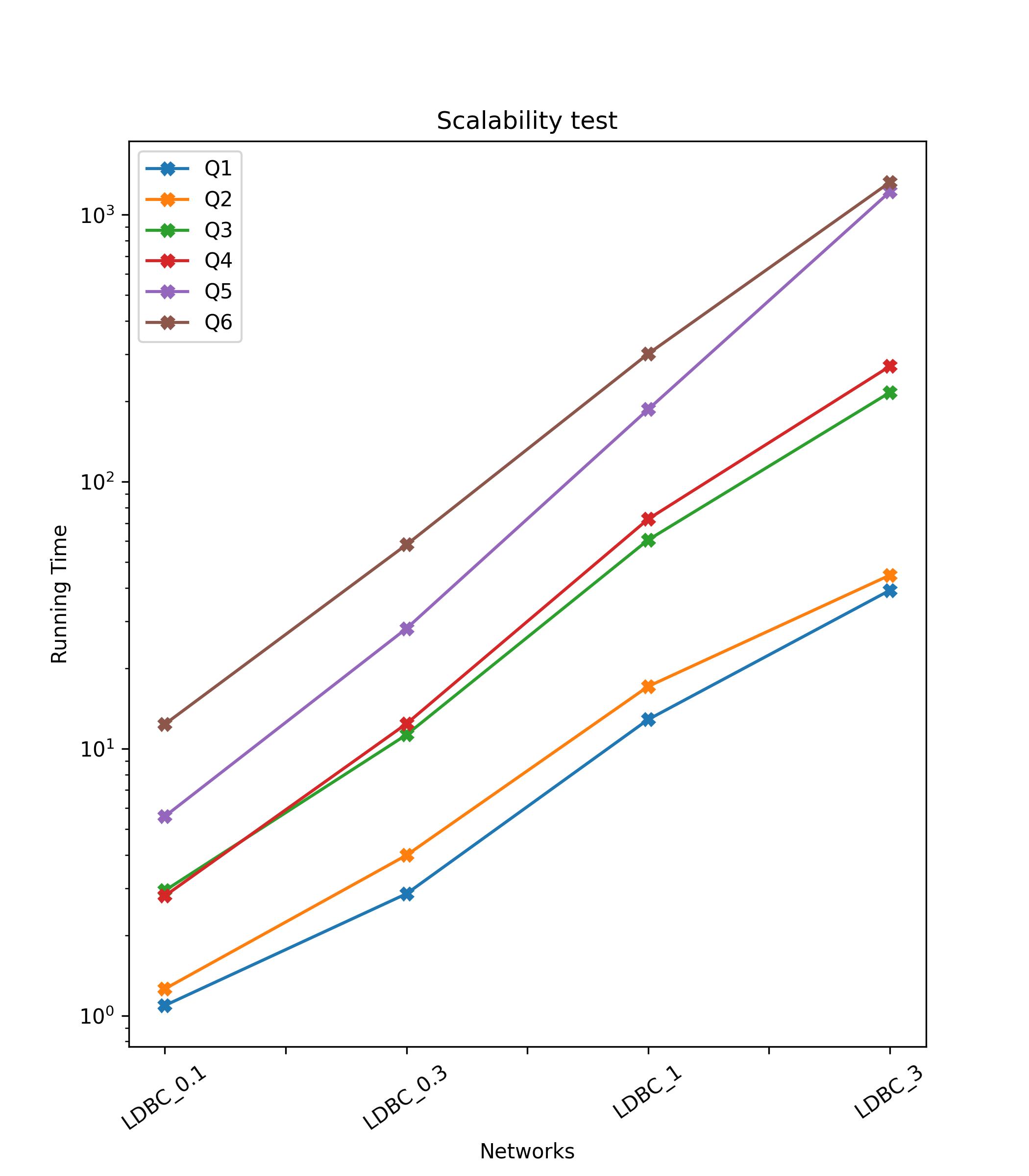}
    \caption{Running times (in seconds) of MultiGraphMatch for the networks listed in Table \ref{tab:real} and the queries depicted in Figure \ref{scalabilityQuery}. 
    For the simplest queries Q1 and Q2, the time scales linearly with network size. For the other queries, there is a power law relationship but with an exponent of 1.5 or less. Please see text for details.}
    \label{scalabilityPlot}
    \Description[Performance on LSQB queries]{Running times (in seconds) of MultiGraphMatch for the networks listed in Table \ref{tab:real} and the queries depicted in Figure \ref{scalabilityQuery}}
\end{figure}

In Figure \ref{scalabilityPlot} we depict the running times of MultiGraphMatch for each query in Figure \ref{scalabilityQuery} and for each social network listed in Table \ref{ldbcNetworks}. 

We further investigated the relationship between the scale factor of the target $s$ and the running time $T$ and found that:
\begin{itemize}
\item For queries Q1 and Q2, $T \sim 15s^{1}$;
\item For queries Q3 and Q4, $T \sim 60s^{1.25}$;
\item For queries Q5 and Q6, $T \sim 240s^{1.5}$.
\end{itemize}
We can conclude that the time rises super-linearly according to a power law relationship of the form $T=as^k$ between $s$ and $T$, with larger $k$ values as the query complexity increases, but the $k$ value is always modest.

In Supplementary Table \ref{ldbcNetworkTimes} we report the time required to index target data structures, whereas Supplementary Table \ref{ldbcQueryTimes} contains, for each experiment, the time required to compute symmetry breaking conditions, processing order of query edges, compatibility domains and matching. The indexing time increases linearly with the size of the target graph, while the time to build domains grows linearly with the query size. Computation of symmetry breaking conditions and ordering of query edges always require negligible time.

\section{Conclusion}
\label{conclusion}
We have proposed both an algorithm and a system to match subgraphs to graphs where both the subgraphs and graphs may have multiple labels on nodes and multiple labeled edges between pairs of nodes. 
The main algorithmic novelties are:

\begin{itemize}
\item 
We use bit matrices on edges to accelerate the construction of compatibility domains, which are sets of target edges that can be matched with query edges.
\item 
We order the query edges according to a new scoring scheme that exploits the cardinalities of the compatibility domains, which reduces the search space of possible solutions.
\item 
We adapt the breaking conditions to the multigraph setting.
\end{itemize}

Experiments performed on both synthetic and real networks show that we are generally faster by a factor ranging from 2 to 10 than state-of-the-art systems. Our improvement is statistically significant (p-value < 0.05) in almost all our experiments. Scalability tests show that MultiGraphMatch scales sub-quadratically with the size of the target, whereas the indexing time and the compatibility domain time rise linearly with the sizes of the target and the query, respectively.

As demonstrated by our ablation tests (see Section \ref{ablationExp}), the most significant improvement of our algorithm over the state of the art comes from the scoring function used to determine the processing order of query edges. This improvement stems from our scoring function's effective combination of node degree and domain cardinalities. Bit matrices contribute to further speed up MultiGraphMatch and represent a good compromise between filtering power and memory consumption.

Future work includes:
\begin{itemize}
\item 
Provide our software to the community version of Neo4j;
\item
Maintain our data structures as the target graph changes; 
\item 
Further optimizations to filtering.
\end{itemize}

%\begin{acks}
%To Robert, for the bagels and explaining CMYK and color spaces.
%\end{acks}

\bibliographystyle{ACM-Reference-Format}
\bibliography{sample-base}


\appendix
\label{appendix}
\section{Supplementary Tables}

\begin{table}[hb]
\renewcommand\thetable{S1}
    \footnotesize
    \centering
    \begin{tabular}{ccccc}
        \toprule        
        \multirow{2}{*}{\bf Network} & \multirow{2}{*}{\bf Comparison} & {\bf Median time} & \multirow{2}{*}{\bf P-value} & {\bf Confidence} \\ 
        & & {\bf difference (secs)} & & {\bf interval} \\
        \midrule        
        \multirow{3}{*}{\textsc{Uniform(2,2)}} & (Neo4J, MGM) & 13.71 & $< 0.0001$ & $[12.02, 18.93]$ \\ 
        & (SuMGra, MGM) & 2.87 & $< 0.0001$ & $[2.46, 4.36]$ \\
        & (Mem, MGM) & 1176.85 & $< 0.0001$ & $[1041.50, 1305.88]$\\ \midrule
        \multirow{3}{*}{\textsc{Uniform(2,10)}} & (Neo4J, MGM) & 5.62 & $< 0.0001$ & $[4.60, 6.67]$ \\ 
        & (SuMGra, MGM) & -0.33 & $0.3351$ & $[-0.36, -0.31]$ \\
        & (Mem, MGM) & 736.76 & $< 0.0001$ & $[566.26, 989.78]$\\ \midrule
        \multirow{3}{*}{\textsc{Uniform(10,2)}} & (Neo4J, MGM) & 2.13 & $< 0.0001$ & $[1.66, 2.72]$ \\ 
        & (SuMGra, MGM) & 0.58 & $< 0.0001$ & $[0.54, 0.65]$ \\
        & (Mem, MGM) & 514.54 & $< 0.0001$ & $[460.38, 1063.16]$\\ \midrule
        \multirow{3}{*}{\textsc{Uniform(10,10)}} & (Neo4J, MGM) & 0.31 & $< 0.0001$ & $[0.27, 0.35]$ \\ 
        & (SuMGra, MGM) & 0.01 & $< 0.0001$ & $[0.01, 0.02]$ \\
        & (Mem, MGM) & 0.97 & $< 0.0001$ & $[0.68, 1.19]$\\ \midrule
        \multirow{3}{*}{\textsc{PowerLaw(2,2)}} & (Neo4J, MGM) & 67.29 & $< 0.0001$ & $[39.02, 96.35]$ \\ 
        & (SuMGra, MGM) & 6.34 & $< 0.0001$ & $[4.89, 11.05]$ \\
        & (Mem, MGM) & 669.46 & $< 0.0001$ & $[399.46, 928.76]$\\ \midrule
        \multirow{3}{*}{\textsc{PowerLaw(2,10)}} & (Neo4J, MGM) & 6.42 & $< 0.0001$ & $[4.74, 8.96]$ \\ 
        & (SuMGra, MGM) & -0.06 & $< 0.0001$ & $[-0.12, -0.01]$ \\
        & (Mem, MGM) & 1336.38 & $< 0.0001$ & $[889.53, 1752.19]$\\ \midrule
        \multirow{3}{*}{\textsc{PowerLaw(10,2)}} & (Neo4J, MGM) & 4.08 & $< 0.0001$ & $[3.13, 5.20]$ \\ 
        & (SuMGra, MGM) & 2.63 & $< 0.0001$ & $[2.14, 3.23]$ \\
        & (Mem, MGM) & 1056.51 & $< 0.0001$ & $[638.95, 1351.53]$\\ \midrule
        \multirow{3}{*}{\textsc{PowerLaw(10,10)}} & (Neo4J, MGM) & 0.41 & $< 0.0001$ & $[0.36, 0.44]$ \\ 
        & (SuMGra, MGM) & 0.04 & $< 0.0001$ & $[0.03, 0.04]$ \\
        & (Mem, MGM) & 2.66 & $< 0.0001$ & $[1.85, 3.81]$\\
        \bottomrule
    \end{tabular}
    \caption{Median, p-value of significance and 90\% confidence interval of the query-by-query running time difference (in seconds) between SuMGra, Neo4J, Memgraph (Mem) and MultiGraphMatch (MGM) on a dataset of queries run on synthetic networks. Uncompleted queries before the timeout were assigned a running time of 1,800 seconds.}
    \label{resSuppSynthetic}
\end{table}

\begin{table}[hb]
%\color{red}
\renewcommand\thetable{S2}
    \footnotesize
    \centering
    \begin{tabular}{ccccc}
        \toprule        
        \multirow{2}{*}{\bf Network} & \multirow{2}{*}{\bf Comparison} & {\bf Median time} & \multirow{2}{*}{\bf P-value} & {\bf Confidence} \\ 
        & & {\bf difference (secs)} & & {\bf interval} \\
        \midrule
        \multirow{6}{*}{\textsc{Uniform(2,10)}} & (MGM-DE, MGM) & 0.41 & $< 0.0001$ & $[0.21, 0.64]$ \\ 
        & (MGM-EL, MGM) & 0.01 & $0.0018$ & $[-0.08, 0.1]$ \\
        & (MGM-DC, MGM) & 0.51 & $< 0.0001$ & $[0.38, 0.63]$ \\ 
        & (MGM-NoBM, MGM) & 2.59 & $< 0.0001$ & $[1.93, 3.02]$ \\
        & (MGM-RO, MGM) & 3.67 & $< 0.0001$ & $[0.46, 6.3]$ \\
        & (MGM-NoBM-RO, MGM) & 6.96 & $< 0.0001$ & $[2.03, 9.88]$ \\ \midrule
        \multirow{6}{*}{\textsc{Uniform(10,2)}} & (MGM-DE, MGM) & 0.35 & $< 0.0001$ & $[0.31, 0.37]$ \\ 
        & (MGM-EL, MGM) & 0.01 & $0.0019$ & $[-0.01, 0.03]$ \\
        & (MGM-DC, MGM) & 0.51 & $0.0049$ & $[0.49, 0.53]$ \\ 
        & (MGM-NoBM, MGM) & 0.57 & $< 0.0001$ & $[0.51, 0.67]$ \\
        & (MGM-RO, MGM) & 0.68 & $< 0.0001$ & $[0.16, 1.08]$ \\
        & (MGM-NoBM-RO, MGM) & 1.32 & $< 0.0001$ & $[0.55, 1.57]$ \\ \midrule
        \multirow{6}{*}{\textsc{PowerLaw(2,10)}} & (MGM-DE, MGM) & 1.93 & $< 0.0001$ & $[0.92, 3.34]$ \\ 
        & (MGM-EL, MGM) & 0.23 & $0.0003$ & $[0.07, 0.45]$ \\
        & (MGM-DC, MGM) & 0.63 & $0.0003$ & $[0.59, 0.66]$ \\ 
        & (MGM-NoBM, MGM) & 2.72 & $< 0.0001$ & $[2.38, 3.37]$ \\
        & (MGM-RO, MGM) & 4.72 & $< 0.0001$ & $[1.15, 15.73]$ \\
        & (MGM-NoBM-RO, MGM) & 9.23 & $< 0.0001$ & $[4.74, 18.24]$ \\ \midrule
        \multirow{6}{*}{\textsc{PowerLaw(10,2)}} & (MGM-DE, MGM) & 0.31 & $< 0.0001$ & $[0.23, 0.58]$ \\ 
        & (MGM-EL, MGM) & 0.04 & $0.102$ & $[0.01, 0.09]$ \\
        & (MGM-DC, MGM) & 0.48 & $0.0022$ & $[0.47, 0.5]$ \\ 
        & (MGM-NoBM, MGM) & 0.66 & $< 0.0001$ & $[0.57, 0.77]$ \\
        & (MGM-RO, MGM) & 0.5 & $< 0.0001$ & $[0.4, 1.69]$ \\
        & (MGM-NoBM-RO, MGM) & 0.98 & $< 0.0001$ & $[0.7, 2.56]$ \\
        \bottomrule
    \end{tabular}
    \caption{\textcolor{black}{Median, p-value of significance and 90\% confidence interval of the query-by-query running time difference (in seconds) between MultiGraphMatch (MGM), MultiGraphMatch with no Bit Matrix (MGM-NoBM), MultiGraphMatch with Random Ordering (MGM-RO), MultiGraphMatch with Ordering based on Minimization of Domain Cardinalities (MGM-DC), MultiGraphMatch with ordering based on minimization of the frequency of Edge Labels (MGM-EL), MultiGraphMatch with ordering based solely on the numerator of the Sc score defined in Section 5.4 (MGM-DE) and MultiGraphMatch with no Bit Matrix and Random Ordering (MGM-NoBM-RO) on the following synthetic networks: Uniform(2,10), Uniform(10,2), PowerLaw(2,10) and PowerLaw(10,2). All heuristics, with one exception, showed statistically significant improvements (p-value < 0.05). Queries that remained uncompleted before the timeout were assigned a runtime of 1,800 seconds.}}
    \label{resSuppAblation}
\end{table}

%\color{black}

\begin{table}
\renewcommand\thetable{S3}
    \footnotesize
    \centering
    \begin{tabular}{ccccc}
        \toprule        
        \multirow{2}{*}{\bf Network} & \multirow{2}{*}{\bf Comparison} & {\bf Median time} & \multirow{2}{*}{\bf P-value} & {\bf Confidence} \\ 
        & & {\bf difference (secs)} & & {\bf interval} \\
        \midrule        
        \multirow{3}{*}{\textsc{IMDB}} & (Neo4J, MGM) & 5.56 & $0.0060$ & $[2.77, 11.26]$ \\ 
        & (SuMGra, MGM) & 4.75 & $0.0479$ & $[3.19, 6.84]$ \\
        & (Mem, MGM) & 0.80 & $0.2914$ & $[0.53, 2.36]$\\ \midrule
        \multirow{3}{*}{\textsc{PANAMA}} & (Neo4J, MGM) & 0.18 & $< 0.0001$ & $[-0.45, 1.30]$ \\ 
        & (SuMGra, MGM) & -0.82 & $< 0.0001$ & $[-1.24, -0.66]$ \\
        & (Mem, MGM) & 6.85 & $< 0.0001$ & $[0.95, 23.58]$\\
        \bottomrule
    \end{tabular}
    \caption{Median, p-value of significance and 90\% confidence interval of the query-by-query running time difference (in seconds) between SuMGra, Neo4J, Memgraph (Mem) and MultiGraphMatch (MGM) on a dataset of queries without WHERE clause extracted from \textsc{IMDB} and \textsc{PANAMA} networks. Uncompleted queries before the timeout were assigned a running time of 1,800 seconds.}
    \label{resSuppRealNoWhere}
\end{table}

\begin{table}
\renewcommand\thetable{S4}
    \footnotesize
    \centering
    \begin{tabular}{ccccc}
        \toprule        
        \multirow{2}{*}{\bf Network} & \multirow{2}{*}{\bf Comparison} & {\bf Median time} & \multirow{2}{*}{\bf P-value} & {\bf Confidence} \\ 
        & & {\bf difference (secs)} & & {\bf interval} \\
        \midrule        
        \multirow{2}{*}{\textsc{IMDB}} & (Neo4J, MGM) & 0.02 & $0.0293$ & $[ 0.01, 0.05]$ \\
        & (Mem, MGM) & -0.06 & $0.0875$ & $[-0.08, -0.04]$\\ \midrule
        \multirow{2}{*}{\textsc{VECTOR}} & (Neo4J, MGM) & 0.06 & $< 0.0001$ & $[0.06, 0.08]$ \\
        & (Mem, MGM) & 1.74 & $< 0.0001$ & $[0.01, 75.24]$\\ \midrule
        \multirow{2}{*}{\textsc{PANAMA}} & (Neo4J, MGM) & 0.10 & $< 0.0001$ & $[0.08, 0.11]$ \\
        & (Mem, MGM) & 6.85 & $< 0.0001$ & $[1.94, 17.54]$\\
        \bottomrule
    \end{tabular}
    \caption{Median, p-value of significance and 90\% confidence interval of the query-by-query running time difference (in seconds) between Neo4J, Memgraph (Mem) and MultiGraphMatch (MGM) on a dataset of queries with WHERE clause extracted from \textsc{IMDB}, \textsc{VECTOR} and \textsc{PANAMA} networks. Uncompleted queries before the timeout were assigned a running time of 1,800 seconds.}
    \label{resSuppRealWhere}
\end{table}
\pagebreak

\begin{table}
\renewcommand\thetable{S5}
    \footnotesize
    \centering
    \begin{tabular}{ccccc}
        \toprule        
        \multirow{2}{*}{\bf Network} & {\bf Network} & {\bf Node} & {\bf Edge} & {\bf Bit matrix} \\ 
        & {\bf reading (secs)} & {\bf indexing (secs)} & {\bf indexing (secs)} & {\bf computation (secs)} \\
        \midrule        
        LDBC\_0.1 & 3.852 & 1.014 & 5.438 & 4.394 \\ \midrule
        LDBC\_0.3 & 8.428 & 2.415 & 19.856 & 16.297 \\ \midrule
        LDBC\_1 & 29.413 & 6.287 & 63.687 & 52.015 \\ \midrule
        LDBC\_3 & 81.510 & 13.815 & 195.217 & 167.249 \\
        \bottomrule
    \end{tabular}
    \caption{Times (in seconds) required to read the 4 LDBC networks used for scalability tests, index their node and edges and build their bit matrices.}
    \label{ldbcNetworkTimes}
\end{table}

\pagebreak
\begin{table}
\renewcommand\thetable{S6}
    \footnotesize
    \centering
    \begin{tabular}{ccccccc}
        \toprule        
        \multirow{2}{*}{\bf Network} & \multirow{2}{*}{\bf Query} & \multirow{2}{*}{\bf Ordering} & {\bf Breaking} & {\bf Compatibility} & \multirow{2}{*}{\bf Matching (secs)} & {\bf Total} \\ 
        & & & {\bf conditions (secs)} & {\bf domains (secs)} & & {\bf time (secs)} \\
        \midrule
        
        \multirow{6}{*}{LDBC\_0.1} & Q1 & 0.006 & 0.003 & 0.765 & 0.317 & 1.091 \\
        & Q2 & 0.008 & 0.004 & 0.699 & 0.548 & 1.259 \\
        & Q3 & 0.008 & 0.003 & 0.642 & 2.287 & 2.940 \\
        & Q4 & 0.008 & 0.003 & 0.101 & 2.696 & 2.808 \\
        & Q5 & 0.007 & 0.004 & 0.056 & 5.510 & 5.577 \\
        & Q6 & 0.007 & 0.003 & 0.723 & 11.618 & 12.351 \\ \midrule

        \multirow{6}{*}{LDBC\_0.3} & Q1 & 0.007 & 0.002 & 1.329 & 1.522 & 2.860 \\
        & Q2 & 0.006 & 0.003 & 0.156 & 3.824 & 3.989 \\
        & Q3 & 0.007 & 0.004 & 0.092 & 11.176 & 11.279 \\
        & Q4 & 0.007 & 0.003 & 1.911 & 10.483 & 12.404 \\
        & Q5 & 0.007 & 0.003 & 1.848 & 26.352 & 28.211 \\
        & Q6 & 0.007 & 0.002 & 2.278 & 53.567 & 55.854 \\ \midrule

        \multirow{6}{*}{LDBC\_1} & Q1 & 0.007 & 0.002 & 5.979 & 6.874 & 12.861 \\
        & Q2 & 0.007 & 0.002 & 6.478 & 10.627 & 17.114 \\
        & Q3 & 0.006 & 0.003 & 5.112 & 55.436 & 60.557 \\
        & Q4 & 0.007 & 0.004 & 1.905 & 70.560 & 72.476 \\
        & Q5 & 0.007 & 0.004 & 0.451 & 186.318 & 186.780 \\
        & Q6 & 0.007 & 0.004 & 7.311 & 294.302 & 301.624 \\ \midrule

        \multirow{6}{*}{LDBC\_3} & Q1 & 0.007 & 0.002 & 20.775 & 18.418 & 39.202\\
        & Q2 & 0.009 & 0.003 & 24.743 & 19.919 & 44.674 \\
        & Q3 & 0.006 & 0.003 & 23.150 & 192.921 & 216.080 \\
        & Q4 & 0.007 & 0.004 & 3.550 & 267.271 & 270.832 \\
        & Q5 & 0.007 & 0.004 & 0.894 & 1219.036 & 1219.941 \\
        & Q6 & 0.007 & 0.004 & 32.141 & 1292.274 & 1324.426 \\
        
        \bottomrule
    \end{tabular}
    \caption{Times (in seconds) required to compute processing order of query edges, symmetry breaking conditions, compatibility domains and matching for each query depicted in Figure \ref{scalabilityQuery} in each target network listed in Table \ref{tab:real}.}
    \label{ldbcQueryTimes}
\end{table}

%\clearpage

%\input{reviewer}

%\clearpage
%\input{reviewer_v2}


\end{document}